\documentclass[floats,floatfix,amssymb,prd,twocolumn,superscriptaddress,nofootinbib,preprintnumbers]{revtex4-2}

\usepackage{subcaption}
\usepackage{ragged2e}
\DeclareCaptionJustification{justified}{\justifying}
\makeatletter
\newcommand{\subsetsim}{\mathrel{\mathpalette\subset@sim\relax}}
\newcommand{\subset@sim}[2]{%
  \vtop{\offinterlineskip\m@th
    \ialign{\hfil##\cr
      $#1\subset$\cr\noalign{\kern0.5pt}\scalebox{0.9}{$#1\sim$}\cr
    }%
  }%
}
\makeatother

\usepackage{amssymb,amsmath,verbatim,mathtools,needspace,enumitem,etoolbox,graphicx,microtype,afterpage,bm}

\usepackage[dvipsnames, usenames]{xcolor}

\definecolor{linkcolor}{rgb}{0.0,0.3,0.5}
\usepackage{booktabs}
\usepackage[unicode, colorlinks=true, linkcolor=linkcolor, citecolor=linkcolor, filecolor=linkcolor,urlcolor=linkcolor, pdfusetitle]{hyperref}
\usepackage[all]{hypcap}
\usepackage[T1]{fontenc}
\usepackage[utf8]{inputenc}
\usepackage{tabularx}
\usepackage{float}
\renewcommand{\arraystretch}{1.4}

\usepackage{multirow}
\usepackage{pifont}
\usepackage{lmodern}

\allowdisplaybreaks
\usepackage{tikz}
\usetikzlibrary{shapes.misc, positioning, arrows.meta}
\usepackage{framed}
\usepackage{hyperref}
\hypersetup{colorlinks, citecolor=bluscuro, linkcolor=black, urlcolor=bluscuro}
\definecolor{rossos}{cmyk}{0,1,1,0.55}
\definecolor{bluscuro}{rgb}{0.15, 0.2, .85}
\definecolor{bluchiaro}{cmyk}{1,.3,0.,0.1}
\definecolor{ForestGreen}{rgb}{0.13, 0.55, 0.13}
\definecolor{TLGreen}{RGB}{50, 164, 49}
\definecolor{TLOrange}{RGB}{231,180,22}
\definecolor{TLRed}{RGB}{204,50,50}

\newcommand{\TLBullet}[1]{\raisebox{-5pt}{\scalebox{0.23}{\begin{tikzpicture}\shadedraw[rounded corners=15pt, top color=gray!84!black,bottom color=black, line width=.6pt] (0,0) rectangle ++(6,2); \ifthenelse{#1=1}{\draw[fill=green,line width=1.pt]  (1,1) circle(.75cm);}{\draw[fill=green!35!black,line width=1.pt]  (1,1) circle(.75cm);}\ifthenelse{#1=2}{\draw[fill=yellow,line width=1.pt]  (3,1) circle(.75cm);}{\draw[fill=yellow!60!black,line width=1.pt]  (3,1) circle(.75cm);}\ifthenelse{#1=3}{\draw[fill=red,line width=1.pt]  (5,1) circle(.75cm);}{\draw[fill=red!50!black,line width=1.pt]  (5,1) circle(.75cm);}\end{tikzpicture}}}}

\newcommand{\aapr}{"Astronomy and Astrophysics Reviews"}

\def\d{{\mathrm{d}}}
\def\PBH{\text{\tiny{PBH}}}

\newcommand{\bs}{\begin{subequations}}
\newcommand{\es}{\end{subequations}}

\newcommand{\be}{\begin{equation}}
\newcommand{\ee}{\end{equation}}
\renewcommand{\d}{{\rm d}}

\newcommand{\llp}{\left [}
\newcommand{\rrp}{\right ]}
\newcommand{\lp}{\left (}
\newcommand{\rp}{\right )}

\def\lsim{\mathrel{\rlap{\lower4pt\hbox{\hskip0.5pt$\sim$}}
    \raise1pt\hbox{$<$}}}         
\def\gsim{\mathrel{\rlap{\lower4pt\hbox{\hskip0.5pt$\sim$}}
    \raise1pt\hbox{$>$}}}         

\usepackage{siunitx}
\DeclareSIUnit \parsec {pc}
\DeclareSIUnit \arcsecondfull {arcsec}
\DeclareSIUnit \year{yr}
\DeclareSIUnit \day{day}
\DeclareSIUnit \hour{hr}
\DeclareSIUnit \radiant{rad}
\DeclareSIUnit \degfull{deg}
\DeclareSIUnit \erg {erg}
\DeclareSIUnit \Lsun {L_\odot}
\DeclareSIUnit \Msun {M_\odot}
\DeclareSIUnit \AstroUnit {au}

\newcommand{\order}[1]{{\cal O}\left(#1\right)}
\newcommand{\divisionsymbol}{\div}
\newcommand{\dd}[1]{{\rm d}#1}
\renewcommand{\Re}{{\rm Re}}

\usepackage{nicefrac}
\usepackage{nicematrix}

\newcommand{\sapienza}{Dipartimento di Fisica, Sapienza Università 
	di Roma, Piazzale Aldo Moro 5, 00185, Roma, Italy}
\newcommand{\infn}{INFN, Sezione di Roma, Piazzale Aldo Moro 2, 00185, Roma, Italy}

\newcommand{\unige}{D\'epartement de Physique Th\'eorique,
Universit\'e de Gen\`eve, 24 quai Ansermet, CH-1211 Gen\`eve 4, Switzerland}
\newcommand{\gwsc}{Gravitational Wave Science Center (GWSC), Universit\'e de Gen\`eve, CH-1211 Geneva, Switzerland}

\newcommand{\unimib}{Dipartimento di Fisica ``G. Occhialini'', Universit\'a degli Studi di Milano-Bicocca,
 Piazza della Scienza 3, 20126 Milano, Italy}

\newcommand{\INFNMi}{INFN, Sezione di Milano-Bicocca, 
Piazza della Scienza 3, 20126 Milano, Italy}

\begin{document}

\title{Searching for Primordial Black Holes with the Einstein Telescope:\\ impact of design and systematics}

\author{Gabriele Franciolini}
\affiliation{\sapienza}
\affiliation{\infn}

\author{Francesco Iacovelli}
\affiliation{\unige}
\affiliation{\gwsc}

\author{Michele Mancarella}
\affiliation{\unimib}
\affiliation{\INFNMi}

\author{Michele Maggiore}
\affiliation{\unige}
\affiliation{\gwsc}

\author{Paolo Pani}
\affiliation{\sapienza}
\affiliation{\infn}

\author{Antonio Riotto}
\affiliation{\unige}
\affiliation{\gwsc}

\date{\today}

\begin{abstract}
Primordial Black Holes (PBHs) have recently attracted much attention as they may explain some of the LIGO/Virgo/KAGRA observations and significantly contribute to the dark matter in our universe. 
The next generation of Gravitational Wave (GW) detectors will have the unique opportunity to set stringent bounds on this putative population of objects. 
Focusing on the Einstein Telescope (ET), in this paper we analyse in detail the impact of systematics and different detector designs on our future capability of observing key quantities that would allow  us to discover and/or constrain a population of PBH mergers. We also perform a population analysis, with a mass and redshift distribution compatible with the current observational bounds. Our results indicate that ET alone can reach an exquisite level of accuracy on the key observables considered, as well as detect up to tens of thousands of PBH binaries per year, but for some key signatures (in particular high--redshift sources) the cryogenic instrument optimised for low frequencies turns out to be crucial, both for the number of observations and the error on the parameters reconstruction. As far as the detector geometry is concerned, we find that a network consisting of two separated L--shaped interferometers of 15 (20)~km arm length, oriented at $45^{\circ}$ with respect to each other performs better than a single triangular shaped instrument of 10 (15)~km arm length, for all the metrics considered.
\end{abstract}

\preprint{ET-0108A-23}

\maketitle

{
  \hypersetup{linkcolor=black}
  \tableofcontents
}
\hypersetup{linkcolor=bluscuro}

\section{Introduction}\label{intro}
The detection of gravitational waves (GWs) 
performed by the LIGO/Virgo/KAGRA (LVK) collaboration opened up a new way for us to observe Nature~\cite{LIGOScientific:2018mvr, LIGOScientific:2020ibl,LIGOScientific:2021djp,LIGOScientific:2021psn}. 
Such discovery brings us information on the compact objects populating our universe and will allow us to learn how they formed. 

A particularly interesting population of objects is represented by Primordial Black Holes (PBHs).
PBHs can form out of the collapse 
of extreme inhomogeneities existing 
during the radiation--dominated era~\cite{Zeldovich:1967lct,Hawking:1974rv,Chapline:1975ojl,Carr:1975qj} and can possess a wide range of masses~\cite{Ivanov:1994pa,GarciaBellido:1996qt,Ivanov:1997ia,Blinnikov:2016bxu}. 
While several constraints were set on their abundance (see~\cite{Carr:2020gox} for a review),
they are still allowed to explain the 
entirety of the dark matter in our universe in
certain mass ranges.
Besides the connection to the dark matter problem, PBHs could be the
 seed of supermassive black holes at high redshift~\cite{2010A&ARv..18..279V,Clesse:2015wea,Serpico:2020ehh}, 
and contribute to a fraction of the BH merger events already discovered by LVK detectors~\cite{Bird:2016dcv,Clesse:2016vqa,Sasaki:2016jop,Eroshenko:2016hmn, Wang:2016ana, Ali-Haimoud:2017rtz, Chen:2018czv,Raidal:2018bbj,Liu:2019rnx, Hutsi:2019hlw, Vaskonen:2019jpv, Gow:2019pok,Wu:2020drm,DeLuca:2020qqa,Hall:2020daa,Bhagwat:2020bzh, DeLuca:2020sae,
Garcia-Bellido:2020pwq,Wong:2020yig,Hutsi:2020sol,Kritos:2020wcl,DeLuca:2021wjr,Deng:2021ezy,Kimura:2021sqz,Franciolini:2021tla,Bavera:2021wmw,Liu:2021jnw,Franciolini:2022tfm,Escriva:2022bwe}
 (see Refs.~\cite{Sasaki:2018dmp,Green:2020jor,Franciolini:2021nvv} for reviews on PBH mergers as GW sources).

It is exciting that future GW detectors,
such as the third--generation~(3G) ground--based interferometers, Einstein Telescope~(ET)~\cite{Hild:2008ng,Punturo:2010zz,Hild:2010id}
and Cosmic Explorer~(CE)~\cite{Reitze:2019iox}, and the future space mission LISA~\cite{2017arXiv170200786A},
could enhance our capabilities to search for GW signatures of PBHs.
ET and CE
are expected to reach much higher sensitivities than current detectors and have a rich science case~\cite{Maggiore:2019uih,Kalogera:2021bya}. 
In particular, for compact binary coalescences, they
will allow for detection rates larger by orders of magnitude compared to current detectors, with  ${\cal O}(10^4-10^5)$ binary black hole (BBH) and binary neutron stars (BNS) detected per year, and 
more accurate measurements for ``golden'' events with high signal--to--noise ratio (SNR)~\cite{Borhanian:2022czq,Iacovelli:2022bbs}.
Furthermore, 3G detectors will allow for entirely new tests on the existence of a population of PBHs~\cite{DeLuca:2021wjr,DeLuca:2021hde,Pujolas:2021yaw,Franciolini:2021xbq,Ng:2022agi}.

We are currently experiencing a crucial stage of the development of 3G detector projects,
in which the community is consolidating the science case,\footnote{For a repository of papers relevant for ET, produced in the context of the activities of the ET Observational Science Board (OSB), see \url{https://www.et-gw.eu/index.php/observational-science-board}.} and PBHs may play an important role in such an endeavor.
Following the recent evolution of our understanding of the physics of PBH mergers (see Ref.~\cite{Franciolini:2021xbq} and references  therein),
the main goal of this paper is to provide 
a systematic analysis of the performance, with respect to PBH searches, of different detector designs under consideration within the ET collaboration. 
In particular, we will highlight which features of the proposed future detector are essential in order to fully exploit its capability of searching for PBHs. The main results of this analysis were included in the official document of the ET Collaboration presenting the full 
study of how the science case of ET depends on different options 
for the detector design ~\cite{Branchesi:2023mws}.

In Sec.~\ref{sec:key} we briefly summarise the 
discriminators which can be used to determine the origin of a merger, based on the observed redshift, masses, spins, eccentricity, and tidal deformability~\cite{Franciolini:2021xbq}.
In Sec.~\ref{sec:ETDes}
we describe the ET designs we considered, following Ref.~\cite{Branchesi:2023mws}, while in  Sec.~\ref{sec:Fisher_general} we introduce our methodology.
Our results focusing on some specific PBH signatures are summarised in Sec.~\ref{sec:results}, where 
we present a detailed comparison between the different designs when key observables are considered. 
In Sec.~\ref{sec:popanalysis} we shift our focus on a comparison of performance adopting population analysis, considering as a benchmark the recent upper bound on the PBH population based on GWTC--3 data, derived in Ref.~\cite{Franciolini:2022tfm}. Finally, we conclude in Sec.~\ref{sec:conclusions} with a summary of our findings and a discussion of future research directions. 

We will consider binaries with individual component masses in the range $\order{10^{-2}\divisionsymbol 10^3}~\si{\Msun}$, 
which are among the main targets for  ET, greatly extending the reach of current LVK searches. 
Throughout this paper we adopt geometrical units ($G=c=1$).
We denote binary components with masses $m_1$ and $m_2$, mass ratio $q=m_2/m_1\leq1$, total mass $M=m_1+m_2$, dimensionless spins $\chi_i$ (with $0\leq \chi_i\leq 1$), and  at redshift $z$.

\section{Key predictions for PBH mergers} \label{sec:key}

In this section we give a summary of the main predictions of PBH models, focusing on mergers in the stellar mass range, which are the prime target of ET searches. 
We refer the reader to  Ref.~\cite{Franciolini:2021xbq} and references therein
for more details.

Throughout this work, we consider 
the standard PBH formation mechanism  and assume that  PBHs are generated from the collapse of sizable cosmological perturbations in the radiation--dominated epoch of the early Universe~\citep{Ivanov:1994pa,GarciaBellido:1996qt,Ivanov:1997ia,Blinnikov:2016bxu}. 
In this scenario, PBHs are predicted to be characterised by small natal spins~\citep{DeLuca:2019buf, Mirbabayi:2019uph}, and are not clustered at high redshift~\citep{Ali-Haimoud:2018dau,Desjacques:2018wuu,Ballesteros:2018swv,MoradinezhadDizgah:2019wjf,Inman:2019wvr,DeLuca:2020jug}. Alternative scenarios, such as the formation 
from the collapse of particles, Q--balls, oscillons, domain walls, and heavy quarks of a confining gauge theory, 
may lead to different predictions for the PBH spin at formation~\cite{Harada:2017fjm,Flores:2021tmc,Dvali:2021byy,Eroshenko:2021sez,DeLuca:2021pls}.
We remark that the impact of accretion (when relevant) onto the mass--spin correlation and the properties of other observables (i.e. redshift distribution, eccentricity, and masses) remain the same as for the standard scenario (see Ref.~\cite{Franciolini:2021xbq} for more details).

\subsection{Redshift}

It can be shown that the merger rate of PBHs is 
dominated by binaries formed in the early Universe via gravitational decoupling from the Hubble flow well before the matter--radiation equality~\cite{Nakamura:1997sm,Ioka:1998nz}. Other dynamical formation channels are possible, i.e., the formation of binaries taking place in present--day halos through gravitational capture or three--body interactions~\cite{Kritos:2022ggc}, but those are subdominant contributions in the parameter space we consider in this work~\cite{Ali-Haimoud:2017rtz,Raidal:2017mfl,Vaskonen:2019jpv,DeLuca:2020jug}.
In contrast to the astrophysical channels, the merger rate density of such population of primordial binary BHs~(BBHs) monotonically increases with redshift as~\citep{Ali-Haimoud:2017rtz,Raidal:2018bbj,DeLuca:2020qqa}
\begin{equation}\label{redevo}
{\cal R}_\text{\tiny PBH} (z) \propto \left[ \frac{ t(z)}{t (z=0)} \right]^{\nicefrac{-34}{37}}\,, 
\end{equation}
extending up to redshifts $z\gtrsim\order{\num{e3}}$. 
We stress that the evolution of the merger rate shown in Eq.~\eqref{redevo} is dictated by how pairs of PBHs decouple from the Hubble flow
before the matter--radiation equality era. 
Therefore,  Eq.~\eqref{redevo} is a robust feature of the PBH channel.

High redshift mergers are difficult to produce within the context of astrophysical channels. In particular, a conservative threshold above which no astrophysical mergers can exist can be set at $z\gtrsim 30$ within standard cosmologies~\cite{Koushiappas:2017kqm}.
It should be stressed that, even though 
the epoch of first star formation is poorly known, theoretical calculations and cosmological simulations suggest 
this should fall below $z\sim 40$~\cite{Schneider:1999us,Schneider:2001bu,Schneider:2003em,Bromm:2005ep,deSouza:2011ea,Koushiappas:2017kqm,Mocz:2019uyd,Liu:2020ufc} (see also Refs.~\cite{Trenti:2009cj} and~\cite{Tornatore:2007ds}, where Pop~III star formation was suggested to start at different epochs).
In addition, a crude estimate for the time delay between Pop~III star formation and the subsequent BBH mergers can be derived from the result of population synthesis models and gives roughly $\order{\SI{10}{\mega\year}}$~\cite{Kinugawa:2014zha,Kinugawa:2015nla,Hartwig:2016nde,Belczynski:2016ieo,Inayoshi:2017mrs,Liu:2020lmi,Liu:2020ufc,Kinugawa:2020ego,Tanikawa:2020cca,Singh:2021zah}. 
Therefore, we consider merger redshifts $z\gtrsim30$ to be smoking guns for primordial binaries~\cite{Koushiappas:2017kqm,DeLuca:2021wjr,Ng:2021sqn,Ng:2022agi,Martinelli:2022elq,Ng:2022vbz}.

\subsection{Masses and spins}\label{PBHmodel_MassSpins}

The distribution of PBH masses $m_\PBH$ is
determined both by the characteristic size and statistical properties of the density perturbations, which are directly inherited from the spectrum of curvature perturbations generated during the inflationary epoch, 
and the PBH threshold for collapse, which retains a dependence on the thermal history~\cite{Jedamzik:1996mr,Byrnes:2018clq,Carr:2019kxo,Franciolini:2022tfm,Escriva:2022bwe,Muscoinprep}.

One of the most important properties of PBHs is that they could populate the subsolar mass range and the supposed astrophysical mass gaps~\cite{Clesse:2020ghq,DeLuca:2020sae,Franciolini:2022tfm}. 
Depending on the PBH mass, a different strategy should be adopted to distinguish PBHs from other astrophysical contaminants. 

In the subsolar range, 
only the primordial scenario 
or beyond standard model physics  (e.g.   dark sector interactions~\cite{Shandera:2018xkn})
can produce binary BHs.
However, white dwarfs, brown dwarfs, or other exotic compact objects~\cite{Cardoso:2019rvt} (e.g. boson stars~\cite{Guo:2019sns}) may also contaminate the signals. These classes of objects could be
distinguished by focusing on the tidal disruption and tidal deformability measurements, which are absent for BBH (see Sec.~\ref{sec:tidal} below).

Just above the solar mass, PBHs can be confused with neutron stars~(NSs). Once again, tidal deformability measurement should be used to break the degeneracy between these two classes of objects. Additionally, solar--mass BHs can form out of NS transmutation in certain particle--dark--matter scenarios~\cite{Bramante:2017ulk,Takhistov:2020vxs,Dasgupta:2020mqg,Giffin:2021kgb,Singh:2022wvw} or asteroid--mass PBH dark matter~\cite{Abramowicz:2022mwb}.
Also, some component BHs in binaries may 
form out of previous NS--NS mergers 
and then pair again to produce a light binary~\cite{Fasano:2020eum}. 
In such a case, 
the second--generation BH formed is expected 
to be spinning~\cite{Hofmann:2016yih}, in contrast with the prediction for the PBH scenario in this mass range, as we shall see.

When focusing on even larger masses $(\gtrsim \SI{3}{\Msun})$, individual PBH mergers can only be distinguished from stellar--origin BHs by using the information on mass--spin correlations.
At high redshift, PBHs are expected to be produced with negligible spin. This is due to a combination of factors. 
Extreme Gaussian perturbations tend to have nearly--spherical shape~\cite{bbks} and the collapse takes place in a radiation--dominated universe, which means that overdensities that collapse to form PBHs evolve fast enough that torques are not able to spin up the collapsing overdensity. As a consequence, the initial dimensionless Kerr parameter $\chi \equiv J/M^2$ (where $J$ and $M$ are the angular momentum and mass of the BH) is expected to be below the percent level~\cite{DeLuca:2019buf,Mirbabayi:2019uph}. 
However, a nonvanishing spin can be acquired by PBHs forming binaries through an efficient phase of accretion~\cite{DeLuca:2020bjf} prior to the reionization epoch.

Accretion was shown to be efficient only for PBHs with masses above $m_\PBH\gtrsim  \order{\SI{10}{\Msun}}$~\cite{DeLuca:2020bjf}.
Therefore, the PBH model is characterised by binary components with negligible spins in the ``light'' portion of the observable mass range of current ground--based detectors. 
At larger masses, 
the PBH model predicts a characteristic correlation between large binary total masses and large values of the spins of their PBH constituents.
This correlation is induced by accretion effects and is subject to the uncertainties of the accretion models (see a detailed discussion on this in Ref.~\cite{DeLuca:2020qqa}). 
In addition, the spin directions of PBHs in binaries are assumed to be independent and randomly distributed on the sphere in Ref.~\cite{DeLuca:2020qqa}. 
We will consider this scenario in the rest of the paper, 
but we warn the reader that details 
of the accretion dynamics remain uncertain.
We refer to Ref.~\cite{Franciolini:2021xbq} for an analytic fit of the relation between the PBH masses and spins as a function of parameterised uncertainties of the accretion efficiency.

\subsection{Eccentricity}\label{sec:ecc}

Another key prediction of the primordial BH model involves the eccentricity $e$ of PBH binaries. While formed with large eccentricity at high redshift, PBH binaries then have enough time to circularize before the GW signal can enter the observation band of current and future detectors.\footnote{
Notice that Refs.~\cite{Cholis:2016kqi,Wang:2021qsu} considered PBH binaries from dynamical assembly in the late--time Universe that retain large eccentricities,  comparable to expectations from the astrophysical dynamical formation scenarios. 
This mechanism, however, provides a subdominant contribution to the overall merger rate in the standard scenario~\cite{Ali-Haimoud:2017rtz,Kritos:2022ggc}.
This situation may be realized with strong PBH clustering suppressing the early--Universe binary merger rate while enhancing the late--time Universe contribution~\cite{Jedamzik:2020ypm,Vaskonen:2019jpv,Trashorras:2020mwn,DeLuca:2020jug}. 
However, we stress that this scenario would require a large value of the PBH abundance ($f_\PBH\simeq 1$), which is in contrast with current PBH constraints in the LVK mass range~\cite{Carr:2020gox} (see also the recent results of Refs.~~\cite{Gorton:2022fyb,Petac:2022rio,DeLuca:2022uvz,Ziparo:2022fnc, Franciolini:2022tfm}).}   Let us clarify this statement.

One can show
that the differential probability distribution 
of the rescaled angular momentum $j\equiv \sqrt{1-e^2}$ parametrically behaves as~\cite{Ali-Haimoud:2017rtz,Kavanagh:2018ggo,Liu:2018ess}
\begin{align}
\label{eq:Pj}
P(j)
= 
4 j \left (\frac{x}{\bar x} \right)^3 
\left [
\frac{\llp 1 +(\sigma_\text{\tiny M}^2/f_\PBH^2) \rrp^{\nicefrac{1}{3}}}
{4 j^2 + \llp 1 + (\sigma_\text{\tiny M}^2/f_\PBH^2)\rrp \left (\frac{x}{\bar x} \right)^6 }
\right]^{\nicefrac{3}{2}}\,,
\end{align}
where $\sigma_\text{\tiny M}\simeq 0.004$ indicates the variance of the Gaussian large--scale density perturbations at matter--radiation equality,
$x$ stands for the comving distance between the two PBH 
forming the binary while $\bar x$ its average distance in the early universe. This distribution
results from the effect of both surrounding PBHs and matter perturbations acting with a torque on the PBH binary during its formation.
Slight refinements to this high redshift distribution was derived in Ref.~\cite{Raidal:2018bbj}, but this does not affect our main argument here. 

The crucial point is the following. Even though the high-redshift distribution of $j$ peaks at small values (i.e. high eccentricities), PBH binaries spend a large portion of the age of the universe inspiralling. Such a phase drives  the eccentricity toward much smaller values. 
We can rearrange the equations that describe the orbital evolution under the effect of GW emission, and define the pericenter distance 
$r_p \equiv  a (1-e)$, with $a$ denoting the semi--major axis, to find~\cite{Maggiore:2007ulw}
\begin{align}
\frac{{\rm d ln} e}{{\rm d ln} r_p} =
 \frac{(1 + e)(304 +121 e^{2})}{192 - 112 e + 168 e^{2} +47 e^{3}}
\approx 1.6 + \order{e}\,.
\label{eq:dedchi}
\end{align}
Let us make a concrete example to gain intuition. For a characteristic PBH binary formed by a narrow PBH population with $f_\PBH = 10^{-3}$ and $m_\PBH = \SI{30}{\Msun}$ and expected to merge within a Hubble time at $z\simeq 0$, 
one obtains a characteristic initial binary pericenter distance to be  $r_p\approx \SI{4e6}{\kilo\meter}$.
We recall that the binary would enter the ET and LVK observable band when its frequency becomes of order of a few Hz  and \SI{10}{\hertz}, respectively. These frequencies correspond roughly to
\begin{align}
{\rm LVK}: &\quad r_p \simeq 22 R_\text{\tiny Sch} \simeq \SI{2e3}{\kilo\meter},
\nonumber \\ \nonumber
{\rm ET}: &\quad 
r_p \simeq 102 R_\text{\tiny Sch} \simeq \SI{9e3}{\kilo\meter},
\end{align}
for $m_\PBH = \SI{30}{\Msun}$. Then, using Eq.~(\ref{eq:dedchi}), one finds that
the observable eccentricity of the orbit 
is reduced below $e\approx 10^{-5}$.

In other words, when the binary enters the detectable frequency band of GW experiments, it is expected to have circularized its orbit to such an extent that the remaining eccentricity is not detectable.
This property can be exploited to distinguish primordial binaries from those produced by astrophysical scenarios.
While isolated formation channels predict small values of $e$ in the observable range of frequencies of 3G detectors, dynamical channels 
 predict a fraction $\order{10\%}$ with $e>0.1$~\cite{Mandel:2018hfr,Benacquista:2011kv,Bae:2013fna,Rodriguez:2017pec,Zevin:2018kzq,Mapelli:2021gyv,Kritos:2022ggc}.
This can be used to distinguish the origin of the latter class of compact objects (see e.g.~\cite{Samsing:2013kua,Nishizawa:2016jji,Breivik:2016ddj,Nishizawa:2016eza,Zevin:2021rtf,Favata:2021vhw}).
Therefore, reaching a sensitivity to $e$ below 
 around $10^{-1}$ indicates the possibility of performing tests on the formation pathway of a binary.

\subsection{Tidal deformability}\label{sec:tidal}

PBHs lighter than a few solar masses 
are easily distinguished from (heavier) stellar--origin BHs.
However, other compact objects might contaminate the solar and subsolar mass range. 
For example, in standard astrophysical scenarios, white dwarfs and NSs are formed with masses above $\approx \SI{0.2}{\Msun}$~\cite{Kilic:2006as} and $\approx \SI{1}{\Msun}$~\cite{Strobel:1999vn,Lattimer:2012nd,Silva:2016myw,Suwa:2018uni}, respectively.

One simple discriminator between PBHs and other compact objects is provided by the Roche radius, ${r_\text{\tiny Roche}}$, defined as the minimum orbital distance below which the secondary object gets tidally disrupted, if it is not a BH. 
The Roche radius is approximately defined as 
$  {r_\text{\tiny Roche}}\sim 1.26 \, r_2 q^{\nicefrac{-1}{3}} $,
in terms of the radius of the secondary object $r_2$ and  the mass ratio $q=m_2/m_1$. 
When ${r_\text{\tiny Roche}}$ is greater than the radius of the innermost stable circular orbit (ISCO), ${r_\text{\tiny Roche}}>r_\text{\tiny ISCO}\sim \order{M}$, the binary gets tidally disrupted before the merger.
The GW signature of such an event would be characterised by an effective cut--off of the GW signal at the frequency corresponding to $r_\text{\tiny Roche}$.
Based on these estimates, one can expect less compact objects, like brown and white dwarfs, to be tidally disrupted well before the contact frequency. 
Other (exotic) compact objects~\cite{Cardoso:2019rvt} (e.g. boson stars~\cite{Liebling:2012fv}) would provide another possible explanation for a (sub)solar compact object.
However, whether or not they get disrupted before the ISCO frequency depends on their compactness.
For example, the vanilla ``mini'' boson star model without self--interactions~\cite{Ruffini:1969qy} with masses around the solar--mass range would be tidally disrupted before the ISCO (see also Ref.~\cite{Franciolini:2021xbq}). 
On the other hand, in the presence of strong scalar self--interactions, boson stars can be as compact as a NS~\cite{Colpi:1986ye,Cardoso:2017cfl}, and
the tidal disruption is not a clear--cut discriminator.

In this paper, to compare the performance of different detector configurations, we will focus on another key discriminator between PBHs and (sub)solar horizonless objects,
that is the absence, in the former case, of tidal deformability contributions to the gravitational waveform. 
The tidal Love numbers are predicted to be vanishing 
for a BH~\cite{Binnington:2009bb,Damour:2009vw,Damour:2009va,Pani:2015hfa,Pani:2015nua,Gurlebeck:2015xpa,Porto:2016zng,
LeTiec:2020spy, Chia:2020yla,LeTiec:2020bos,Hui:2020xxx,Charalambous:2021kcz,Charalambous:2021mea,Pereniguez:2021xcj}, 
whereas they are generically nonzero and model--dependent for any other compact object~\cite{Cardoso:2017cfl}.
The tidal Love numbers enter the GW phase 
starting at 5 post--Newtonian (PN) order,
with the $5$PN and $6$PN contributions being linearly related to the 
dominant quadrupolar tidal Love number, $\lambda_2^{(i)} = 2m_{i}^5 k_2^{(i)}/3$, of the $i$\textsuperscript{th} body~\cite{Flanagan:2007ix,Vines:2011ud}.
In the Newtonian approximation, the tidal Love number of an object is (see e.g.~\cite{PoissonWill})
\begin{equation}
    k_2^{(i)} \sim \order{0.01-0.1} \left(\frac{r_i}{m_i}\right)^5\,,
\end{equation}
where the precise value of the dimensionless prefactor depends on the nature of the object (i.e.,  in the case of a NS, on the equation of state). Therefore, less compact objects 
are characterised by a larger tidal deformability and would 
be more easily distinguished from a (P)BH.\footnote{
A further discriminator could be tidal heating corrections to the waveform in the case of BHs, which is due to dissipation at the event horizon~\cite{Alvi:2001mx}
while remaining negligible for other compact objects~\cite{Cardoso:2019rvt,Maselli:2017cmm}. However, tidal heating is generically subdominant with respect to the tidal deformability correction presented above. Therefore, we do not include it in our analysis.
}

\section{ET detector designs and networks} \label{sec:ETDes}

In this section, we provide a short summary of the different designs and detector networks considered in this study. For more details, we direct the reader to the official ET document presenting the full scientific part of the study of how the science output changes with the detector design, performed
in ref.~\cite{Branchesi:2023mws}. 

We will consider three main geometries for ET: the triangular geometry ($\bigtriangleup$), consisting of three nested detectors with \ang{60} arm aperture, a network of 2 L--shaped detectors (i.e. with \ang{90} arm aperture) with parallel arms (2L-\ang{0}) and a network of 2 L--shaped detectors with misaligned arms (2L-\ang{45}).\footnote{Notice that the notion of parallel and misaligned has to be intended with respect to the local East rather than with reference to the great circle connecting the detectors. See Sec. 2 and App.~A of Ref.~\cite{Branchesi:2023mws} for further discussion regarding this choice.} 
Among the configurations consisting of two well separated L--shaped detectors, the parallel one has the advantage of maximizing the sensitivity to stochastic backgrounds and can result in a higher SNR for some of the observed sources, while the misaligned one has no blind spots and is optimal in terms of accuracy on the reconstruction of the sky position and distance of the detected events. Regarding the sites of the detectors, there are two main candidates by the time of writing: the Sos Enattos site in Sardinia, Italy, and the Meuse--Rhine  Euroregion across Belgium, Germany, and the Netherlands.\footnote{For the Sardinia site the chosen example coordinates are [lat=\ang{40;31;00}, long=\ang{9;25;00}], while for the 
Meuse--Rhine site [lat=\ang{50;43;23}, long=\ang{5;55;14}].} 
When considering a single triangular instrument, we choose for definiteness the Sardinia site (no major difference is expected if choosing the Meuse--Rhine site), while for the 2L configurations, we place one of the detectors in Sardinia and one in Meuse--Rhine.

On top of studying different geometries, as done in Ref.~\cite{Branchesi:2023mws} we will consider different arm lengths for the detectors, namely \SI{10}{\kilo\meter} and \SI{15}{\kilo\meter} for the triangular geometry and \SI{15}{\kilo\meter} and \SI{20}{\kilo\meter} for the 2L geometries, and two different scenarios: one in which the so--called `xylophone' design is employed, with each detector actually consisting of two interferometers, one optimised for low--frequency (LF) sensitivity (operating at cryogenic temperature) and the other for high--frequency (HF) sensitivity (we will refer to this configuration as HFLF-Cryo), and a pessimistic scenario in which only the high--frequency optimised instrument will be operating (we will refer to this configuration as HF). 
Considering a sort of `intermediate' scenario, in which the detectors exploit the xylophone design with the LF instrument operating at room temperature, the results would be intermediate among the ones we will show for the HFLF-Cryo and HF scenarios. 

\begin{figure}[!t]
	\centering
	\includegraphics[width=0.49\textwidth]{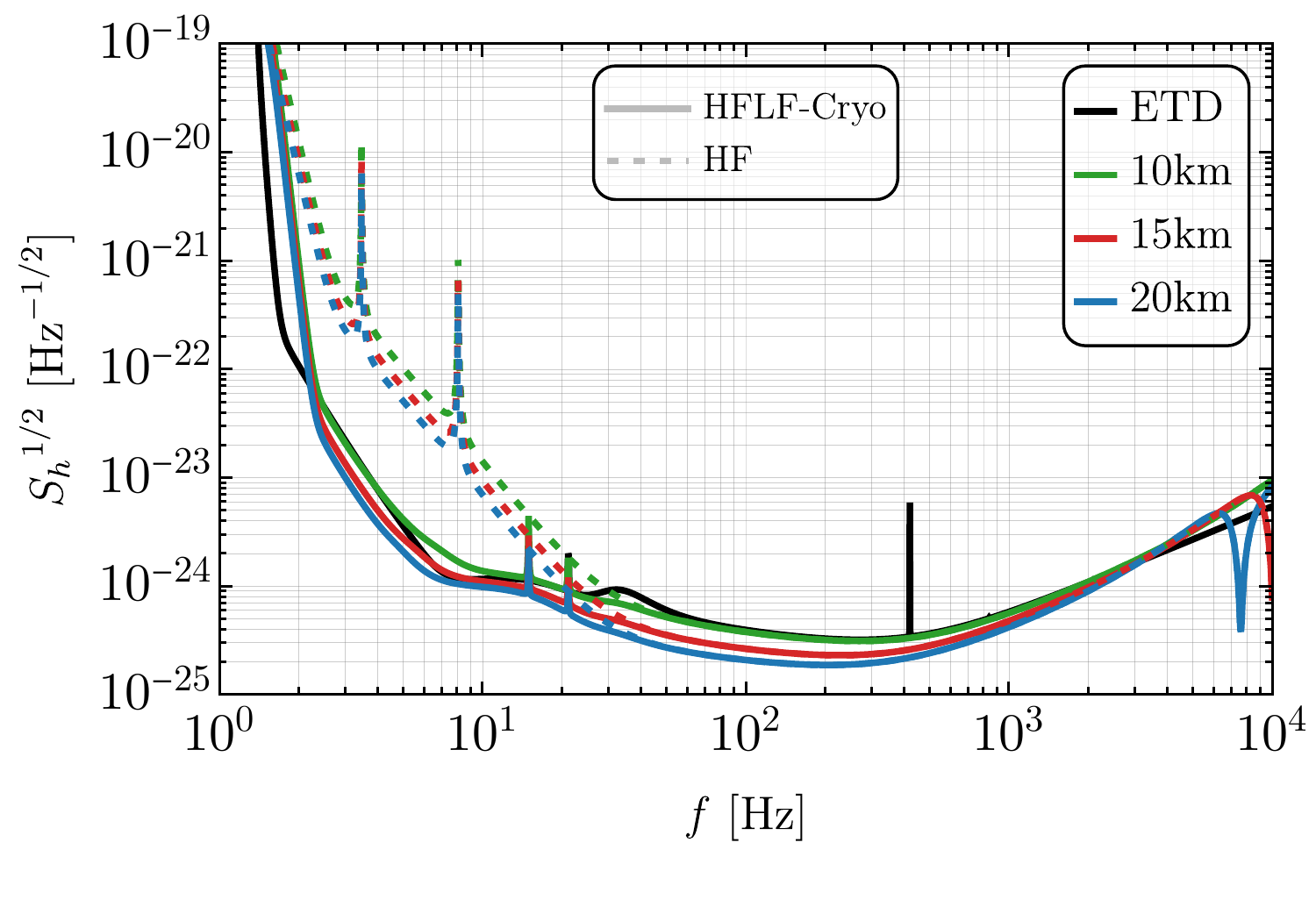}
	\caption{Amplitude spectral densities for the various configurations considered in this work. 
	Solid and dashed lines corresponds to 
	cryogenic low--frequency and high--frequency--only designs, respectively. We also show the ET-D design in black for comparison. }
\label{fig:PSDs}
\end{figure}

The different amplitude spectral densities (ASDs) employed in the various cases are reported in Fig.~\ref{fig:PSDs}, where we further show the ET-D sensitivity curve for comparison.\footnote{The ET-D sensitivity curve is available at \url{https://apps.et-gw.eu/tds/?content=3&r=14065}.} These curves have been provided by the ET Instrument Science Board (ISB) and are an update of the ET-D curve, reflecting  more detailed data on the technology used in the ET-D design.\footnote{The curves are available at \url{https://apps.et-gw.eu/tds/ql/?c=16492}, and are obtained using the \textsc{PyGWINC} package 
\cite{2020ascl.soft07020R}.} It should be stressed that in this phase of the ET development the sensitivity curves are still in evolution, and will continue to evolve, so these ASDs must be taken as examples, within a range of other  possibilities currently under study.

In Sec.~\ref{sec:popanalysis} we will further compare with the results that can be obtained using a single L--shaped instrument of \SI{20}{\kilo\meter} (placed for definiteness in the Meuse--Rhine site) as well as for ET (in its different geometries) as part of a 3G network, both with a single CE detector with \SI{40}{\kilo\meter} arms and two CE instruments with \SI{40}{\kilo\meter} and \SI{20}{\kilo\meter} arms, which is the preferred CE design at the moment~\citep{Evans:2021gyd, Srivastava:2022slt}.\footnote{For the CE detectors, we choose the same coordinates reported in Tab. III of~\cite{Borhanian:2020ypi} for the U.S. sites, locating the \SI{40}{\kilo\meter} instrument in Idaho [lat=\ang{43;49;36}, long=\ang{-112;49;30}] and the \SI{20}{\kilo\meter} instrument in New Mexico [lat=\ang{33;9;36}, long=\ang{-106;28;49}]. The latest CE sensitivity curves are available at \url{https://dcc.cosmicexplorer.org/CE-T2000017/public}.} We will also consider the current network of second generation (2G) detectors, namely the two LIGO detectors, Virgo and KAGRA, also including LIGO India, (LVKI network) with ASDs representative of the best forecasted sensitivity for the upcoming fifth observing run (O5)~\cite{KAGRA:2020rdx}.\footnote{The ASD files are available at \url{https://dcc.ligo.org/LIGO-T2000012/public}.}

\section{Signal to noise ratio and Fisher matrix: notation and setup} \label{sec:Fisher_general}

Our analysis will focus on two complementary aspects of GW detections, which are our capability of detecting a particular source and the precision one can achieve at determining the source parameters. 
As we shall see, both aspects are crucial to chart future capabilities to search for the elusive population of PBHs. 
We will set ourselves in a simplified setting. In order to determine the detectability of a source, we will set a threshold for detection on the signal--to--noise ratio (SNR).
Regarding the measurability of source properties, the Fisher information matrix (FIM) is  often used to assess the parameter estimation capabilities of GW detectors (see, for example, Refs.~\cite{PhysRevD.46.5236,PhysRevD.47.2198,Cutler:1994ys,Poisson:1995ef,Berti:2004bd,Ajith:2009fz,Cardoso:2017cfl}, as well as Refs.~\cite{Vallisneri:2007ev,Rodriguez:2013mla} for discussions of the limitations of this approach).
While realistic performance may deviate from what is estimated on the basis of SNR and FIM, 
we expect such metrics to provide a consistent framework to compare various detector designs, following the aim of this project. 
Here, in order to compute the expected SNR and the FIM,
we adopt the publicly available code \textsc{GWFAST} developed by some of the authors 
~\cite{Iacovelli:2022mbg,Iacovelli:2022bbs}.\footnote{The \textsc{GWFAST} code is available at \url{https://github.com/CosmoStatGW/gwfast}.}\textsuperscript{,}\footnote{
 Other parameter estimation codes, tuned toward 3G detectors, have been developed recently, in particular \textsc{GWBENCH}~\cite{Borhanian:2020ypi,Borhanian:2022czq},   \textsc{GWFISH}~\cite{Dupletsa:2022wke}, \textsc{TiDoFM}
\cite{Chan:2018csa,Li:2021mbo}
and the code used in Ref.~\cite{Pieroni:2022bbh}.
Results from these codes were cross checked and are consistent with each other~\cite{Branchesi:2023mws, Iacovelli:2022bbs}.
} 

The output $s(t)$ of a general GW interferometer can be written as the sum of the GW signal $h (t,\vec \theta)$ and the detector noise $n(t)$, assumed to be Gaussian and stationary with zero mean. The posterior distribution for the hyperparameters $\bm \theta$ can be approximated by
\be\label{pos_dist_F}
p ({\bm \theta}| s) \propto \pi ({\bm \theta}) \exp\left\{-\frac{1}{2}(h ({\bm \theta}) - s|h ({\bm \theta}) - s)\right\}\,,
\ee
in terms of the prior distribution $\pi ({\bm \theta})$. Here we have introduced the inner product
\be\label{innprod}
(g|h) = 4\Re\int_{f_\text{\tiny min}}^{f_\text{\tiny max}} \frac{\tilde{g}^* (f) \tilde{h} (f) }{S_n(f)} \,\dd{f}\,.
\ee
In Eq.~\eqref{innprod}, the tilde denotes a temporal Fourier transform, $S_n(f)$ is the detector noise power spectral density (PSD), while $f_\text{\tiny min}$ and $f_\text{\tiny max}$ are the detector minimum and maximum frequency of integration, respectively. 
One can therefore define the SNR as 
\begin{equation}
    {\rm SNR}\equiv\sqrt{(h|h)}\,.
\end{equation}
 
In the limit of large SNR, one can perform a Taylor expansion of Eq.~\eqref{pos_dist_F} and get 
\be\label{eq:FisherPost}
p ({\bm \theta}| s) \propto \pi ({\bm \theta}) \exp\left\{-\frac{1}{2}\Gamma_{ab} \Delta \theta^a \Delta \theta^b\right\}\,,
\ee
where $\Delta {\bm \theta} = {\bm \theta}_\text{\tiny p} - {\bm \theta}$, and we neglected noise--dependent factors restricting to the statistical uncertainty. In this set--up, ${\bm \theta}_\text{\tiny p}$ represents the posterior mean values, that also coincides with the true binary parameter ${\bm \theta}_p={\bm \theta}_\text{\tiny true}$. 
Also, we have introduced the Fisher matrix 
\be
\Gamma_{ab} \equiv \lp \frac{\partial h}{\partial \theta^a} \bigg | \frac{\partial h}{\partial \theta^b} \rp_{{\bm \theta} = {\bm \theta}_\text{\tiny p}}\,.
\ee
The errors on the hyperparameters (that represent a measure of the width of the posterior distribution) are, therefore, given by $\sigma_a = \sqrt{\Sigma^{aa}}$, where $\Sigma^{ab} = \lp \Gamma^{-1}\rp^{ab}$ is the covariance matrix.\footnote{For the results in Sec.~\ref{sec:results}, we impose physical priors on the waveform parameters to overcome limitations in the Fisher matrix related to bad conditioning. In particular, we draw samples from the distribution in Eq.~\eqref{eq:FisherPost}, use rejection sampling to enforce the prior, and compute the covariance of the remaining samples. The procedure is detailed in App.~C of~\cite{Iacovelli:2022bbs}.}

\begin{figure*}[!th]
	\centering
	\includegraphics[width=0.75 \textwidth]{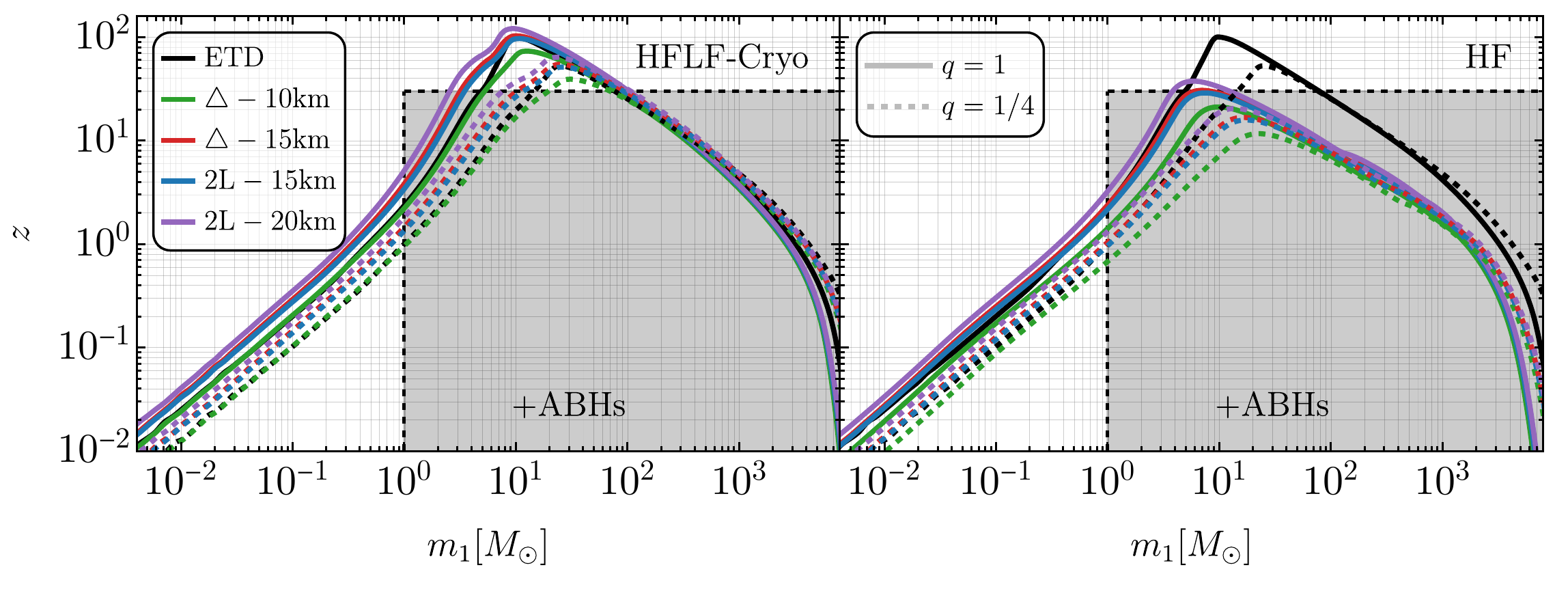}
	\caption{ Horizon redshift $z$ as a function of primary mass for $q = 1$ (solid lines) and $q = 1/4$ (dashed lines). 
	The left panel corresponds to the HFLF-Cryo configuration and the right panel to the HF configuration while different colors indicate various geometries. We also show the ET-D design for comparison, while the gray area delimits the subsolar $m_1<\si{\Msun}$ and $z>30$ regions, which are PBH smoking--gun signatures.
	The binaries are assumed to be circular ($e_0 = 0$) and BHs spins are assumed to be negligible ($\chi_{1} = \chi_{2} = 0$), as predicted by the PBH scenario. The shaded region marked `+ABH' denotes (approximately) the  part of the mass--redshift plane to which astrophysical BBHs are confined.
    As a consequence of our choices ($\iota=\ang{0}$, and optimal location), aligned and misaligned 2L configurations result in the same performance for this kind of estimation, we thus report only the 2L-\ang{45} case.}
\label{fig:horizon}
\end{figure*}

In the most general case, restricting to quasi--circular orbits, the binary parameters are 15 for a BBH  (see e.g.~\cite{Maggiore:2007ulw})
\begin{equation}
{\bm \theta} _\text{\tiny BBH}
= \{{\cal M}_c, \eta, d_L, \theta, \phi, \iota, \psi, t_c, \Phi_c, \chi_{1,c}, \chi_{2,c} \}\,,    
\end{equation}
where $c = \{x,y,z\}$, 
extended to 18 in case of one allows for tidal deformability of either components and orbital eccentricity
\begin{equation}
{\bm \theta} _\Lambda
    = \{\Lambda_1, \Lambda_2\},
\qquad 
{\bm \theta}_e= \{e_0\}\,.
\end{equation}
In particular,  ${\cal M}_c$ denotes the detector--frame chirp mass, $\eta$ the symmetric mass ratio, $d_L$ the luminosity distance to the source, $\theta$ and $\phi$ are the sky position coordinates, defined as $\theta=\pi/2-\delta$ and $\phi=\alpha$ (with $\alpha$ and $\delta$ right ascension and declination, respectively), $\iota$ the inclination angle of the binary with respect to the line of sight, $\psi$ the polarisation angle, $t_c$ the time of coalescence, $\Phi_c$ the phase at coalescence, $\chi_{i,c}$ the dimensionless spin of the object $i=\{1,2\}$ along the axis $c = \{x,y,z\}$ and $\Lambda_i$ the dimensionless tidal deformability of the object $i$, which is zero for a BH. 
A more convenient parametrisation of tidal effects, which is the one we will use in the analysis, is given in terms of the quantities~\citep{Wade:2014vqa} 
\begin{equation}
    \begin{aligned}
        \tilde{\Lambda} &= \dfrac{8}{13} \left[(1+7\eta-31\eta^2)(\Lambda_1 + \Lambda_2) \right.\\
        & \left. + \sqrt{1-4\eta}(1+9\eta-11\eta^2)(\Lambda_1 - \Lambda_2)\right]\,,\\
        \delta\tilde{\Lambda} &= \dfrac{1}{2} \left[\sqrt{1-4\eta} \left(1-\dfrac{13272}{1319}\eta + \dfrac{8944}{1319}\eta^2\right)(\Lambda_1 + \Lambda_2) \right.\\
        & \left.+ \left(1 - \dfrac{15910}{1319}\eta + \dfrac{32850}{1319}\eta^2 + \dfrac{3380}{1319}\eta^3\right)(\Lambda_1 - \Lambda_2)\right]\,;
    \end{aligned}
\end{equation}
which has the advantage that $\tilde{\Lambda}$ is the combination entering the waveform at 5PN, and is thus better constrained than $\Lambda_1$ and $\Lambda_2$ separately, while $\delta\tilde{\Lambda}$ first enters at 6PN.
Including also a small eccentricity in the orbit, there is one more parameter to consider, $e_0$, which denotes the eccentricity at a given reference frequency, which we fix at $f_{e_{0}} = \SI{10}{\hertz}$.

For the various analyses in Sec.~\ref{sec:results} we adopt different state--of--the--art waveform models.
In Sec.~\ref{subsec:high-z_res}  and \ref{subsec:subsolar_res}, 
we employ the inspiral--merger--ringdown (IMR) model \textsc{IMRPhenomHM}, which includes the contribution of the higher--order harmonics $(l,m) = (2,1),\, (3,2), \, (3,3)\, (4,3)$ and $(4,4)$ on top of the $(2,2)$ quadrupole~\citep{London:2017bcn,Kalaghatgi:2019log}; 
this neglects possible precession of sources, but it is justified in our case as high--redshift or light events are characterised by negligible spins in the PBH scenario.\footnote{
Even though we are injecting spinless binaries in the section under consideration, adopting a waveform model which only allows for aligned spins reduces the number of parameters in the Fisher. 
We do not expect this to significantly impact our results, and it corresponds to parameter estimations that adopts physically motivated 
narrow spin priors.} 
In Sec.~\ref{subsec:ecc_res} we use the inspiral--only model \textsc{TaylorF2}~\citep{Buonanno:2009zt, Ajith:2011ec, Mishra:2016whh} with the extension to include a small orbital eccentricity presented in~\cite{Moore:2016qxz};\footnote{When using \textsc{TaylorF2}, we cut the inspiral phase at the ISCO frequency of a remnant Schwarzschild BH with mass $\approx M_\text{\tiny tot}$. Extending the Fisher analysis to higher frequencies would improve measurement errors, but would also push the validity of the waveform model to its limit.} and in Sec.~\ref{subsec:lambda_res} we use in a broad mass range \textsc{TaylorF2} with tidal terms at 5PN and 6PN~\citep{Wade:2014vqa} and spin--induced quadrupole moment~\citep{Yagi:2016bkt} and the full inspiral--merger model \textsc{IMRPhenomD\_NRTidalv2}~\citep{Husa:2015iqa, Khan:2015jqa, Dietrich:2019kaq} in its tuning range.
Finally, in Sec.~\ref{subsec:spin_mass_res} and in the population analysis of Sec.~\ref{sec:popanalysis} instead, 
we adopt the full IMR model \textsc{IMRPhenomXPHM}, 
which includes both the contribution of higher--order harmonics and precessing spins~\citep{Pratten:2020fqn, Pratten:2020ceb}. 
This choice is dictated again by the expectation of the PBH scenario, which does not predict tidal deformability effects, nor sizeable eccentricity for the detectable events.

\section{Results} \label{sec:results}

In this section we report our results, maintaining the structure described in Sec.~\ref{sec:key}. We will also provide a summary of our results in the conclusions. 

Throughout the work, we will report results for the various detector geometry and network with different colors, while splitting the two configurations HFLF-Cryo and HF into left and right panels, respectively
(see e.g. Fig.~\ref{fig:horizon}). We also draw dense grid--lines to guide the eye in the comparison of the various designs.

\begin{figure*}[!ht]
	\centering
	\includegraphics[width=0.6\textwidth]{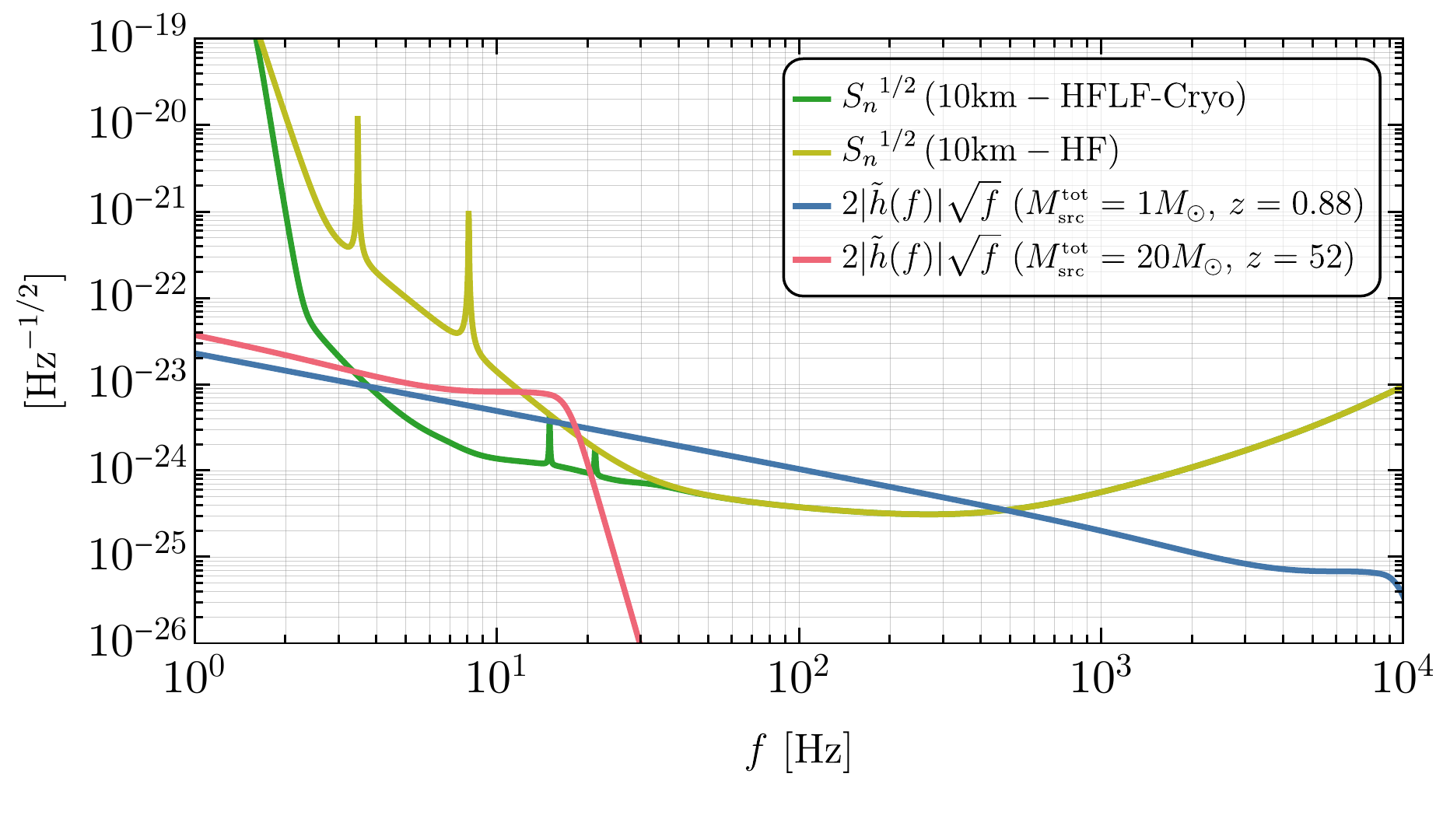}
	\caption{Comparison between the amplitude of a PBH binary signal and the ET detector sensitivity curves shown in green (10km-HFLF-Cryo) or yellow (10km-HF). 
	We consider a distant merger with source frame mass $M_\text{\tiny src}^\text{\tiny tot} = \SI{20}{\Msun}$ at $z = 52$, reported in red. Also, we show in blue a subsolar merger with $M_\text{\tiny src}^\text{\tiny tot} = \SI{1}{\Msun}$ at $z = 0.88$. 
	In both cases, we assume $q=1$, optimal orientation of the binary and the redshift is fixed in order to have SNR=9 in the $\bigtriangleup$-10km-HFLF-Cryo configuration.}
\label{fig:LFvsPBHs_signal}
\end{figure*}

\subsection{High--redshift measurements}\label{subsec:high-z_res}

High redshift mergers are a prime target of 3G detector searches. In Fig.~\ref{fig:horizon}, we show the horizon redshift (defined as the maximum distance at which a source could be detected, assuming optimal alignment)
in the whole spectrum of source frame masses available to ET, i.e. $M \approx [10^{-2}, 10^4]~\si{\Msun}$.
We assume equal mass systems, negligible spins and eccentricity, as expected for primordial BBHs, and set a threshold ${\rm SNR}\geq8$ for detection. 

Two main trends limit the reach of ET. At small masses, GW amplitudes are reduced, 
and only sufficiently nearby sources can be loud enough to be detected. 
At higher masses, on the other side of the spectrum, a strong limitation comes from the 
intrinsically small frequencies spanned by the GW signal, which can exit the frequency spectrum to which ET is most sensitive.
In particular, the detector frame ISCO frequency scales as 
\begin{equation}
    f_\text{\tiny ISCO}^{\text{\tiny det}}
    \approx 
    \SI{4.4}{\kilo\hertz} 
    \lp \frac{\si{\Msun}}{m_1^\text{\tiny src}+m_2^\text{\tiny src}}\rp 
    \lp \frac{1}{1+z}\rp\,,
\end{equation}
from which we clearly see that signals heavier than 
$M\gtrsim \SI{e2}{\Msun}$ at redshift $z\gtrsim 10$ would end up at small, unreachable frequencies in the ET frame. This already indicates the importance of the LF implementation.
To gain further intuition on the relevance of the LF instrument when searching high--redshift mergers, in Fig.~\ref{fig:LFvsPBHs_signal} we compare the signal amplitude for a merger with source--frame mass $M_\text{\tiny src}^\text{\tiny tot} = \SI{20}{\Msun}$ located close to the horizon (i.e. having SNR $=9$ in the $\bigtriangleup$-10km-HFLF-Cryo configuration). Due to the large source redshift, the detector frame mass is redshifted by a large factor, and the majority of SNR can only accumulate in the LF portion of the sensitivity curve.

\begin{figure*}[!t]
	\centering
	\includegraphics[width=0.75 \textwidth]{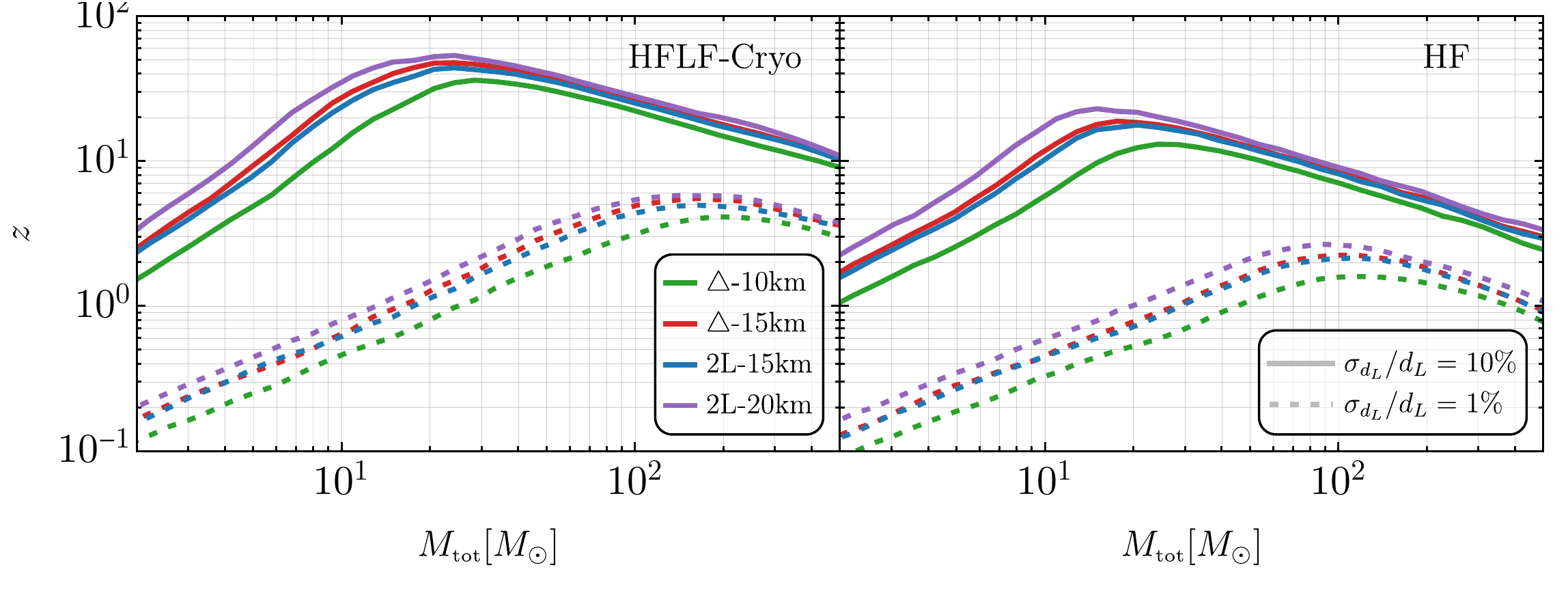}
	\caption{Contour lines of constant relative error as a function of source frame total mass $M_\text{\tiny tot}$ and source redshift $z$. Different colors indicate different configurations, while the left panel assumes HFLF-Cryo ASD, and the right panel HF ASD. The solid (dashed) lines correspond to  $\sigma_{d_L}/{d_L} = 10\%$ (1\%). We assume equal masses $m_1 = m_2$, vanishing spins $\chi_1 = \chi_2 = 0$, and optimal orientation of the binary. 
    We report only the 2L-\ang{45} case as in Fig.~\ref{fig:horizon}.
    }
\label{fig:distance_error_highz}
\end{figure*}

Looking at Fig.~\ref{fig:horizon}, we see that the reach of ET configurations can only be larger than $z\gtrsim 30$ (for a narrow window of masses $m_1\approx [5, 10^2]~\si{\Msun}$ assuming $q\approx 1$)
if the low--frequency instrument is implemented. In fact, the cryogenic implementation, 
with enhanced sensitivity at $f<\SI{10}{\hertz}$, 
provides the best design in terms of the number of distant detectable mergers.

The smoking--gun high redshift signature can  be employed (provided a sufficient precision is achieved when estimating the source redshift), when both individual event analyses are concerned~\cite{Ng:2021sqn,Franciolini:2021xbq,Ng:2022vbz}, as well as population studies~\cite{Ng:2022agi,Martinelli:2022elq}.
Based on a FIM metric, 
we compare the performance of different configurations in Fig.~\ref{fig:distance_error_highz}, which reports the maximum source redshift above which a relative precision on $d_L$ better than $10\%$ and $1\%$ cannot be achieved for equal--mass non--spinning systems.\footnote{For a complementary visualisation of the performance of different configurations one may adopt the inference horizon introduced in Ref.~\cite{Mancarella:2023ehn}.}

As observed in terms of the horizon for detection, the relative error $\sigma_{d_L}/{d_L}$ is strongly impacted by the performance of the detector configurations at low frequencies and follows the general trend observed in the behaviour of the horizon, see Fig.~\ref{fig:horizon}.
Longer detector arms lead to larger SNR and reduced uncertainties. At the same time, we observe comparable performance between the $\bigtriangleup$-15km and 
2L-15km, while 2L-20km remain the best configuration we tested. 
Again, the absence of the low--frequency instrument would lead to a drastic reduction in performance, which completely prevents the exploitation of the high--redshift signature to discover PBHs. 

The same conclusion is supported by the population analysis presented in the following section (see, in particular, Tab.~\ref{tab:allConf}).

\subsection{Subsolar merger measurements}\label{subsec:subsolar_res}

In this subsection, we focus on the second smoking--gun signature of PBHs, which is represented by the subsolar mass window.
ET will dramatically extend the reach of current technology when searching for mergers with at least one component below the Chandrasekhar limit~\cite{DeLuca:2021hde,Pujolas:2021yaw}.

In Fig.~\ref{fig:horizon} we show the detection horizon also
in the subsolar mass range, i.e. the range of masses where only PBHs, white dwarfs, brown dwarfs, or exotic compact objects~\cite{Cardoso:2019rvt} can appear.
For such masses  a large portion of the SNR is built up in the HF range of the sensitivity curve, so
the horizons are only slightly dependent on the LF instrument. 
This is supported by Fig.~\ref{fig:horizon}, 
where, for subsolar masses, no significant difference in horizon redshift is observed between various configurations when removing the LF instrument. 
This can be understood by noticing that the GW signal of subsolar mergers can only be observed in its inspiral phase,
which crosses the whole frequency range accessible to ET and the SNR is mainly dominated by the range of frequencies where the experiment is most sensitive, which remain fixed regardless of the presence of the LF instrument (see Fig.~\ref{fig:PSDs}). 
This is also shown in Fig.~\ref{fig:LFvsPBHs_signal}, where we compare the signal amplitude for a subsolar merger with source frame mass $M_\text{\tiny src}^\text{\tiny tot} = \SI{1}{\Msun}$ located close to the horizon (i.e. having SNR $=9$ in the $\bigtriangleup$-10km-HFLF-Cryo configuration).
One note of caution is in order here. While our considerations are based on SNR detection threshold, the actual matched--filtering search for GW signals is  considerably more complex. In particular, in the case of light events, no merger signal is available and the SNR is accumulated throughout the inspiral, spanning over the sensitivity band. As such, our considerations here 
are based on a simplified metric, which anyway allows for a consistent comparison of designs.

\begin{figure*}[!t]
	\centering
	\includegraphics[width=0.75\textwidth]{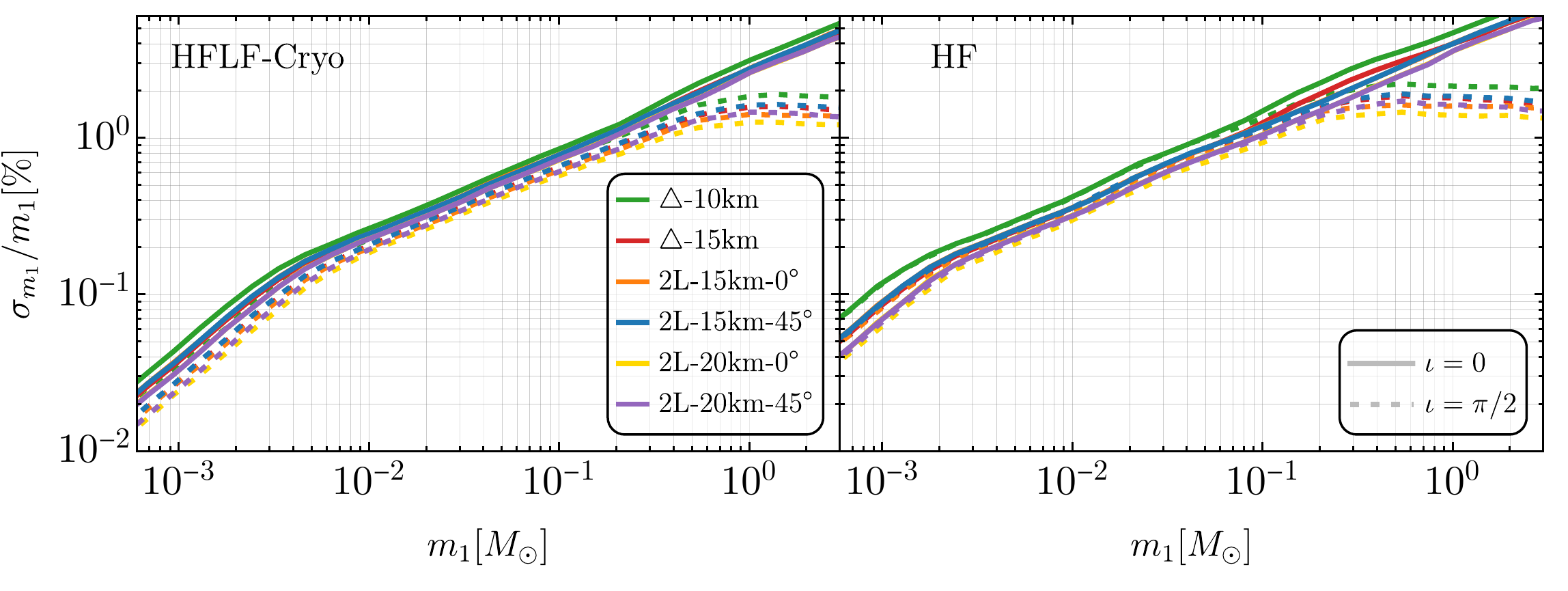}
	\includegraphics[width=0.75\textwidth]{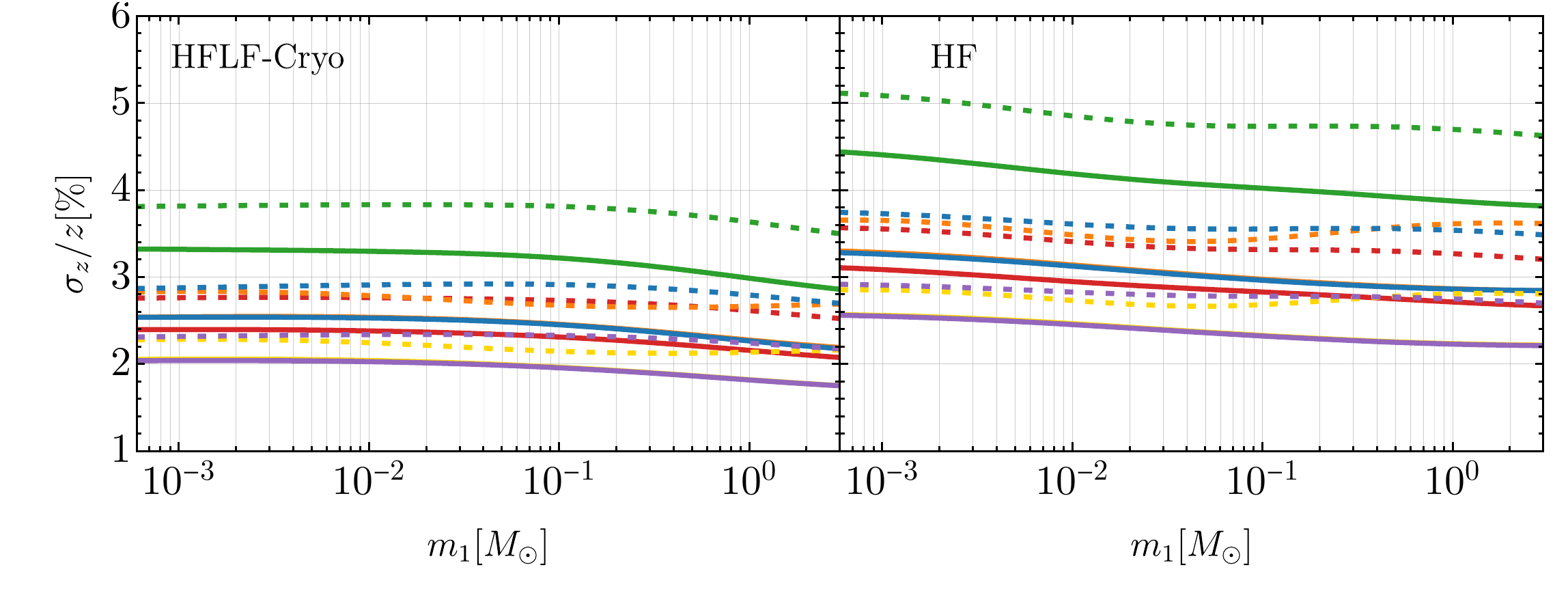}
	\caption{ 
	Relative error on the primary source frame mass (top) and source redshift (bottom) as a function of $m_1$ in the subsolar range. 
	We consider a source located at a distance such that the SNR = 30 in the $\bigtriangleup\text{-10km}$ HFLF-Cryo configuration. }
\label{fig:subsolar}
\end{figure*}

\begin{figure*}[!t]
	\centering
	\includegraphics[width=0.75\textwidth]{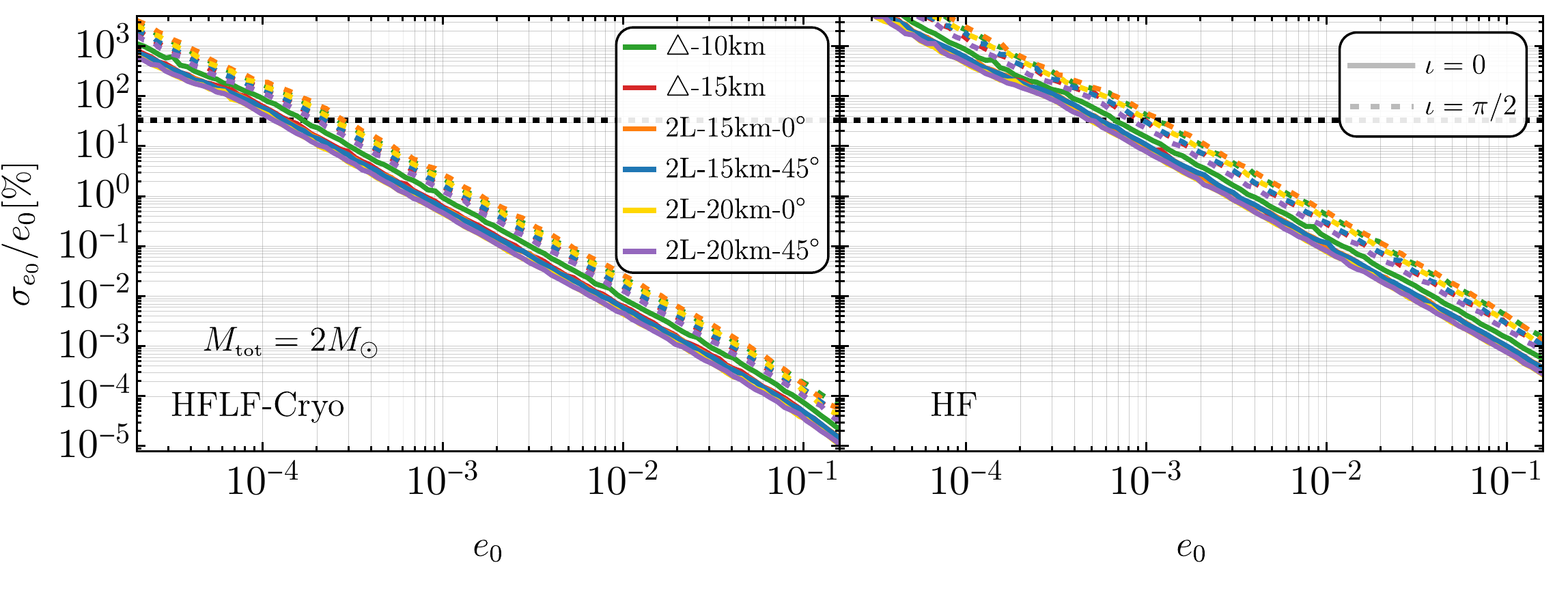}
	\includegraphics[width=0.75\textwidth]{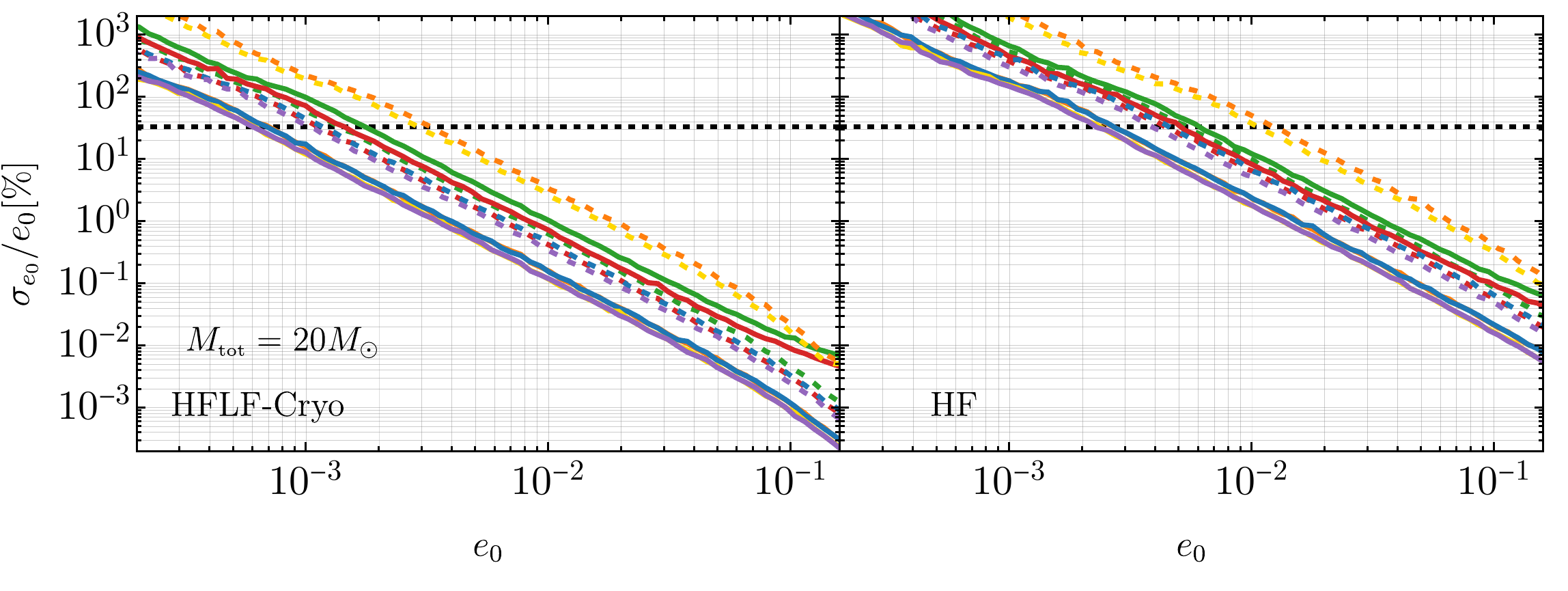}
	\caption{ 
	Relative error on the eccentricity $e_0$ at $f_{e_{0}} = \SI{10}{\hertz}$ as a function of its injected value for 
	a binary with total source frame mass of 
	$M_\text{\tiny tot} = \SI{2}{\Msun}$ and $d_L=\SI{100}{\mega\parsec}$ in the upper panel, 
	while $M_\text{\tiny tot} = \SI{20}{\Msun}$ and $d_L=\SI{500}{\mega\parsec}$
	in the lower panel.
  One is able to exclude negligible eccentricity (at 3-$\sigma$ level) below the horizontal gray dashed lines.}
\label{fig:eccentricity}
\end{figure*}

In Fig.~\ref{fig:subsolar} we compare the performance of different ET configurations when measuring the source frame mass and distance of subsolar PBH mergers.
To perform a proper comparison between different designs, 
for each mass $m_1$, we assume a source located at the same distance, which is arbitrarily fixed in order to have SNR$=30$ in the $\bigtriangleup\text{-10km}$ HFLF-Cryo configuration. 
BHs spins are assumed to be negligible ($\chi_{1} = \chi_{2} = 0$), as predicted by the PBH scenario.
We adopt the \textsc{IMRPhenomHM} waveform model and test both face--on ($\iota = 0$) and edge--on ($\iota = \pi/2$) binaries.

When measuring the primary mass, we observe a large precision improvement on $\sigma_{m_1}$ going towards lighter masses. This is due to the larger number of cycles that the binary spends inspiraling within the observable band. In all subsolar mass range, a sub--percent precision is achieved for sources with at least SNR$\gtrsim 30$, and 
the improvements in the LF sensitivity result in a minor gain of the precision on $m_1$ and $z$.

The conclusion is that, for subsolar mass BHs,   both  the detection horizon  and  the accuracy on the source frame mass and redshift reconstruction do not depend crucially on the LF  sensitivity, and
missing the LF instrument does not jeopardise the ET capability to detect subsolar--mass signatures. 
On the other hand, the length of the arms, which leads to a larger SNR, is the main responsible for the different performance of the various designs.

\subsection{Eccentricity measurements}\label{subsec:ecc_res}

Another key prediction of the PBH model involves the eccentricity $e$. 
As discussed in Sec.~\ref{sec:ecc}, while formed with large eccentricity at high redshift, 
PBH binaries circularize before the GW signal can enter the observation band of current and future detectors~\cite{Franciolini:2021xbq}.
Therefore, PBH binary candidates must have small eccentricities.

To test the sensitivity of the various configurations to the measurement of a possible orbital eccentricity, we use the \textsc{TaylorF2} inspiral--only waveform model~\cite{Buonanno:2009zt, Ajith:2011ec, Mishra:2016whh} with the extension presented in~\cite{Moore:2016qxz} to account for a small eccentricity in the orbit. Using as reference values for the total source--frame mass \SI{2}{\Msun} and \SI{20}{\Msun}, for each detector configuration, we compute the relative error that can be attained on equal mass, non--spinning systems, as a function of the eccentricity
$e_0$ defined at $f_{e_{0}} = \SI{10}{\hertz}$. The distance to the source is fixed to $d_L=\SI{100}{\mega\parsec}$ for the case 
$M_\text{\tiny tot} = \SI{2}{\Msun}$ and to $d_L=\SI{500}{\mega\parsec}$ for $M_\text{\tiny tot} = \SI{20}{\Msun}$.

The results for the relative errors attainable on $e_0$ are reported in Fig.~\ref{fig:eccentricity}. 
We notice, in particular, the great relevance of the LF instrument to perform eccentricity measurements, 
which may result in a reduction of $\sigma_{e_0}$ of up to one order of magnitude. This can be traced to the fact that eccentricity gives larger effects during the inspiral (i.e. at low frequencies) since, when going closer to merger, a binary system tends to circularize, with $e_0\propto f_{\rm GW}^{-\nicefrac{19}{18}}$.
The range of values of $e_0$ shown on the horizontal axis of Fig.~\ref{fig:eccentricity} corresponds to eccentricities that are too large for a PBH binary, that are expected to have $e_0\sim 10^{-6}$ when they reach  $f=10$~Hz \cite{Franciolini:2021xbq}. 
As a consequence, in this range, when the relative error on $e_0$ 
for a given detection is sufficiently small, 
e.g. below the horizontal dashed line in the figure 
(which corresponds to $\Delta e_0/e_0 = 0.33$), 
one would be able to exclude that this event has  a primordial origin.

Finally, for fixed SNR, heavier mergers result in a smaller number of cycles within the ET observable frequency band, and a corresponding reduction of precision (i.e. larger $\sigma_{e_0}$). As far as the comparison between configurations is concerned, we checked that analogous conclusions can be drawn from simulating detections with larger masses.
A smaller improvement is brought by longer detector arms, which can amount to slightly larger SNR.

\subsection{Tidal deformability measurements}\label{subsec:lambda_res}

Measurements of tidal deformability represent another crucial indicator of the compact nature of a merger.
In particular, focusing on the light portion of the mass spectrum accessible to ET, tidal effects are an important discriminant between astrophysical stars (of various nature) and light PBHs. 

\begin{figure*}[!t]
	\centering
	\includegraphics[width=0.75\textwidth]{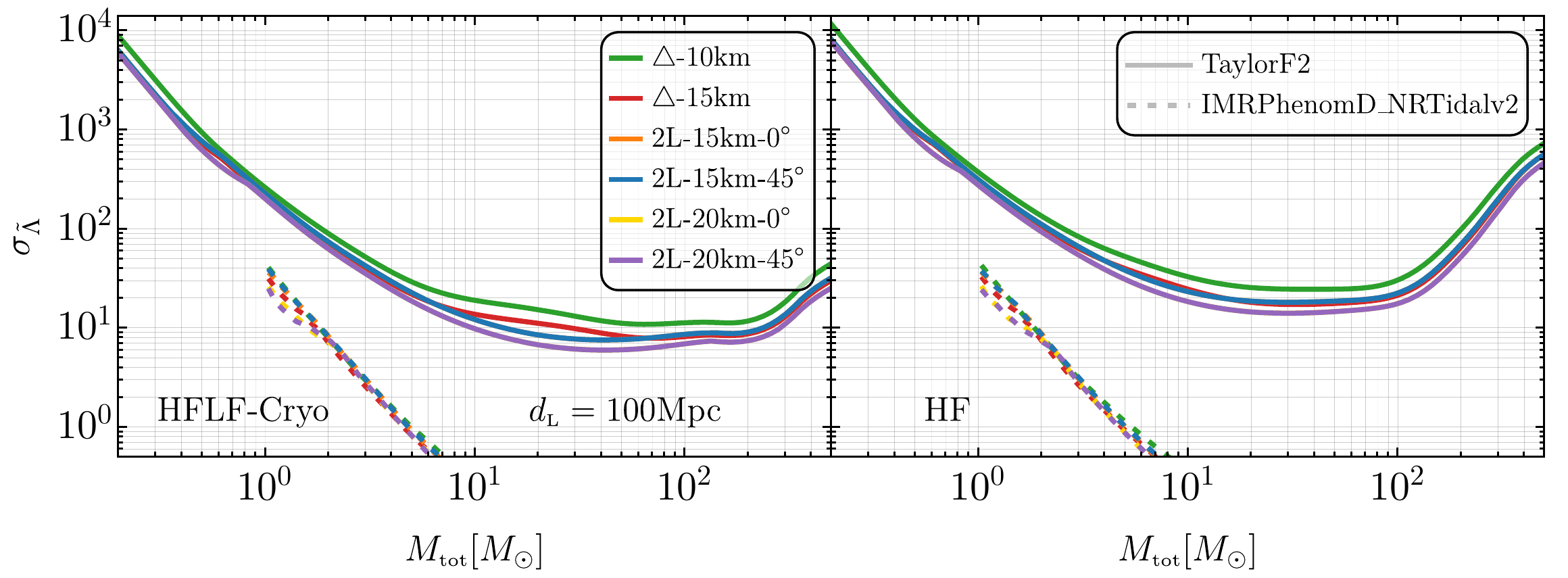}
	\caption{
	Absolute (1-$\sigma$) uncertainty on the tidal deformability $\tilde \Lambda$
	as a function of total mass $M_\text{\tiny tot}$ in the source frame for the various configurations.
	We assume negligible deformability of the source, non--spinning equal--mass binaries, and an optimally oriented source at a distance of $\SI{100}{\mega\parsec}$. 
 The aligned 2L configurations are not visible as they result in comparable performance as for the misaligned case, due to our choice of selecting an optimally oriented source. 
 We show as solid lines the results obtained over a broad mass range with the \textsc{TaylorF2} inspiral--only model, while for comparison we also show the results obtained with the full inspiral--merger model \textsc{IMRPhenomD\_NRTidalv2} in its range of validity.}
\label{fig:err_tidal_deformability}
\end{figure*}

In Fig.~\ref{fig:err_tidal_deformability}
we show the absolute error on the tidal parameter $\tilde \Lambda$ as a function of the source frame total mass, for non--spinning, equal--mass, and optimally oriented binaries,
located at \SI{100}{\mega\parsec} 
from the detector, with negligible tidal deformability. 
The parameters we include in the FIM are 
\begin{equation}
    \{{\cal M}_c, \eta, d_L, \theta, \phi, \iota, \psi, t_c, \Phi_c, \chi_{1,z}, \chi_{2,z}, \tilde{\Lambda}, \delta \tilde{\Lambda}\}\,.
\end{equation} 
We adopt the \textsc{TaylorF2} waveform model with the inclusion of tidal effects.

We also checked that, at the level of the comparison between different configurations, assuming the \textsc{IMRPhenomD\_NRTidalv2} waveform approximant gives analogous results. However, in the range of masses of its validity, the latter provides a smaller uncertainty on $\tilde \Lambda$, as can be seen from the dashed lines in Fig.~\ref{fig:err_tidal_deformability}. This is due to the fact that \textsc{TaylorF2} only describes the inspiral part of the signal, while \textsc{IMRPhenomD\_NRTidalv2} is a full inspiral--merger model, and includes a more complete tidal phase (recall that the tidal terms we use in \textsc{TaylorF2} only appear at 5PN and 6PN order) as well as tidal terms in the amplitude (which are not present in \textsc{TaylorF2}). We further verified that, as a consequence of the more complete tidal phase and amplitude, even if cutting the \textsc{IMRPhenomD\_NRTidalv2} at $2\,f_{\rm ISCO}$, the uncertainty on $\tilde \Lambda$ is still more than 10 times smaller than the one estimated using \textsc{TaylorF2}.

Overall the performance of the various configurations for low--mass systems is similar, with more differences appearing for higher masses, where the 2L configurations perform better compared to the triangle. We also notice the improvement thanks to the LF instrument for masses $M_\text{\tiny tot}\gtrsim \SI{1}{\Msun}$, and in particular for $M_\text{\tiny tot}\gtrsim \SI{100}{\Msun}$: this can be traced to the fact that a signal can stay in band for a much higher number of cycles thanks to the LF sensitivity, resulting in smaller errors on the parameters.
Here again, we see that longer arms lead to slightly reduced uncertainties due to the larger expected SNR.

\begin{figure*}[!t]
	\centering
\includegraphics[width=0.75 \textwidth]{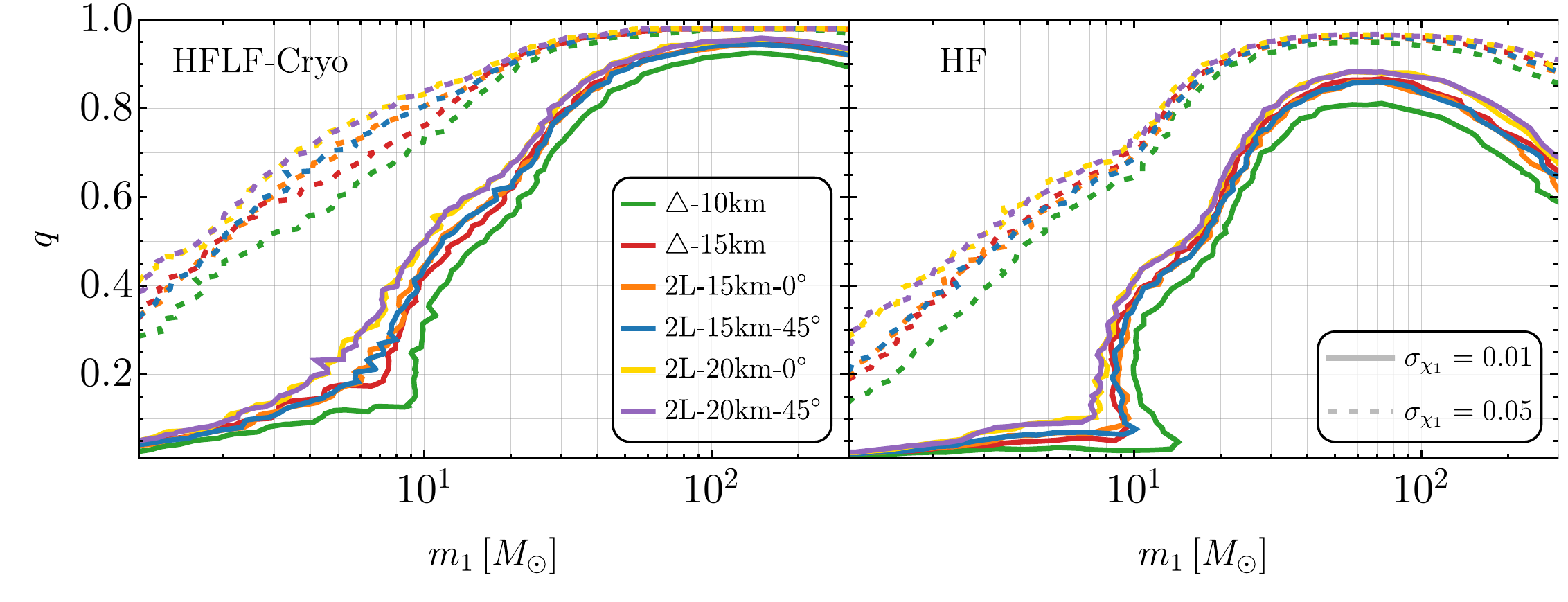}
\includegraphics[width=0.75 \textwidth]{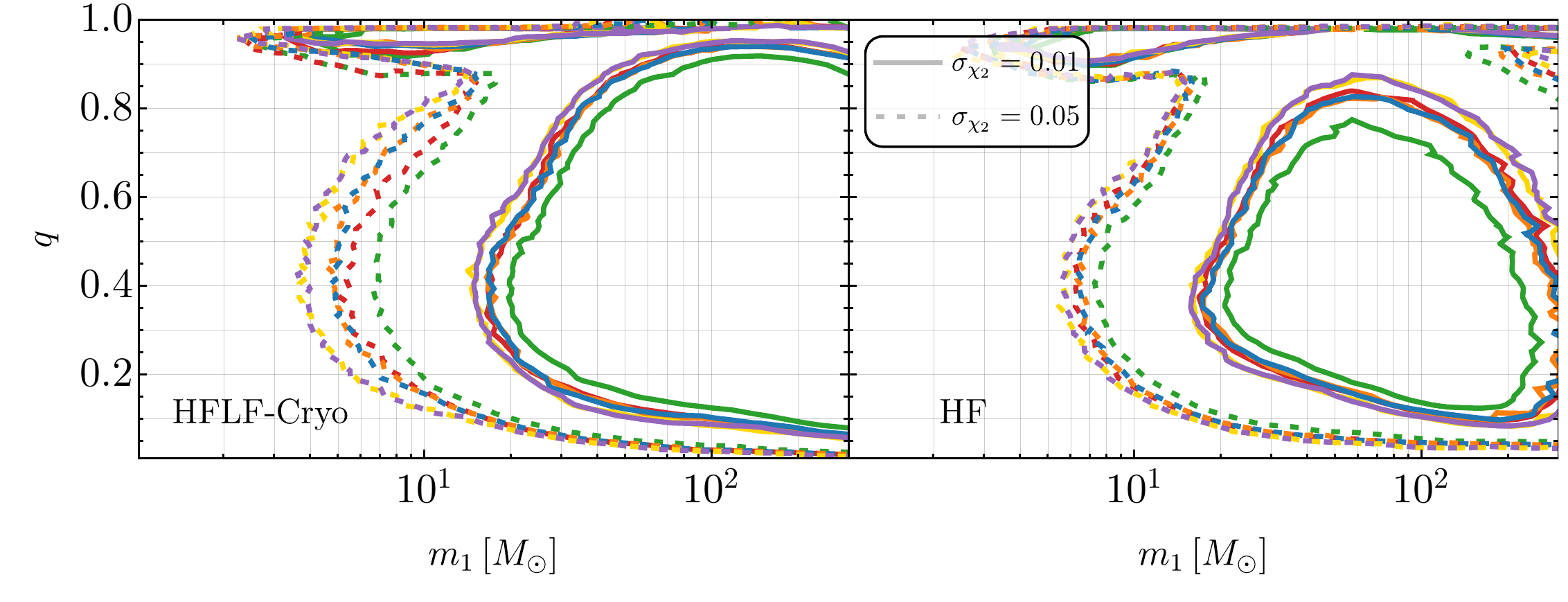}
	\caption{
 Absolute errors on both primary (top panel) and secondary (bottom panel) spin magnitude. 
 We assume a merger located at $d_{L} = \SI{500}{\mega\parsec}$ with an optimal orientation. We also inject negligible BH spins, a value which is compatible with the primordial scenario and inefficient accretion.
 The errors decrease on the right/below of each contour line, mainly driven by the larger SNR.
 }
\label{fig:mass_spin}
\end{figure*}

\subsection{Spin measurements}\label{subsec:spin_mass_res}

As discussed in Sec.~\ref{subsec:spin_mass_res}, 
in the standard formation scenario, PBH binaries composed of individual PBHs lighter than $m_\text{\tiny PBH}\simeq \SI{10}{\Msun}$ 
retain their initial small spins, 
as accretion is always not efficient enough
to spin up individual components. 
Thus, measuring a non--vanishing spin 
for a sub--\SI{10}{\Msun} object would be in tension with a standard primordial origin. 
Provided modelling of PBH accretion will be accurate enough, the mass--spin relation can (at least in principle) be compared with a single event to test its consistency with a primordial origin for larger masses.  
In order to compare the performance of various designs, in this section we forecast the accuracy with which ET could measure spins as a function of both primary mass and mass ratio, for selected individual events. 
In  section~\ref{sec:popanalysis} we will instead show the results of a comparison based on a population analysis, which yields analogous trends. 

In Fig.~\ref{fig:mass_spin} we show the parameter space in the $(m_1,q)$ plane in which an absolute
$1\sigma$ error of $\sigma_{\chi_{i,z}} =1\%$ ($\sigma_{\chi_{i,z}} =5\%$) can be achieved. 
In this analysis, we consider vanishing spins and provide absolute errors on the spin magnitude along the angular momentum axis $\hat z$. 
We consider a merger at a fixed distance of $d_L = 500{\rm Mpc}$. Therefore, moving towards the right--hand side (upper side) of the plot, which means larger total masses, corresponds to considering higher values of SNR.
This drives an increased precision in all configurations we considered. 
The performance of the detectors would, of course, improve for closer sources.

Under the assumption of aligned spins, which we adopt to perform the FIM analysis, 
in the limit $q=1$ it is difficult to make independent measurements of the individual component spins. 
This is because in such a limit, the dominant terms in the waveform model are completely determined by $\chi_{s} \equiv (1/2) (\chi_{1,z}+\chi_{2,z})$, 
and the derivative with respect to the antisymmetric spin $\chi_{a} \equiv (1/2)(\chi_{1,z}-\chi_{2,z})$ in the Fisher information matrix is suppressed by the mass difference between the two component of the binary.  
Therefore, in this limit, the large uncertainty on $\chi_{a}$ jeopardise our ability to constrain the individual spin magnitudes $\chi_{1,z}$ and $\chi_{2,z}$.
In the opposite limit, $q \ll 1$, the error on the secondary spin magnitude increases again, due to the suppressed sensitivity on the lighter component of the binary.

As one can see, no significant difference in performance between the various configurations is observed. The main difference is brought by the 
length of the arms, and the consequent larger SNR achieved. 
The region of parameter space where sub--percent accuracy on $\chi_{i,z}$ is achieved is systematically smaller for the $\bigtriangleup$-10km, while it increases with 2L configurations and longer arms, due to the slightly larger SNR.

\section{Population analysis}\label{sec:popanalysis}

In this section we analyse in detail the detection prospects of the various ET configurations with a population approach. 
In general, forecasts based on PBH populations must be based on underlying assumptions, given no definite evidence for their existence is available to date. 
In this section, we are going to assume a PBH population that saturates the upper bounds provided by current GWTC--3 GW data, as recently derived in Ref.~\cite{Franciolini:2022tfm}. 
This should be viewed as an optimistic maximum contribution from PBHs to future 3G detections and serve as a benchmark for us to compare the performance of different ET configurations. 

\subsection{PBH merger rate distribution}

In the standard formation scenario that we assume throughout, the PBH merger rate at low redshift is dominated by binaries that gravitationally decouple from the Hubble flow before the matter--radiation equality~\cite{Nakamura:1997sm,Ioka:1998nz}.
We compute the differential volumetric PBH merger rate density following~Refs.~\cite{Raidal:2018bbj, Vaskonen:2019jpv,DeLuca:2020jug,DeLuca:2020qqa}, as
\begin{align}
\label{eq:diffaccrate}
 \frac{\d  {\cal R}_\PBH}{\d m_1 \d m_2}
& = 
\frac{\num{1.6e6}}{\si{\cubic\giga\parsec\year}} 
f_\PBH^{\nicefrac{53}{37}} 
\lp \frac{t(z)}{t_0} \rp^{\nicefrac{-34}{37}}  
\lp \frac{M}{\si{\Msun}} \rp^{\nicefrac{-32}{37}}  
 \nonumber \\
& \times
\eta^{\nicefrac{-34}{37}}
S(M, f_\PBH, \psi, z)
\psi(m_1) \psi (m_2)\,,
\end{align}
where $\eta = m_1 m_2/M^2$, 
$t_0 = \SI{13.7}{\giga\year}$ is the current age of the Universe, and $\psi(m_i)$ is the PBH mass distribution.

\begin{figure*}[!t]
	\centering
	\includegraphics[width=0.49\textwidth]{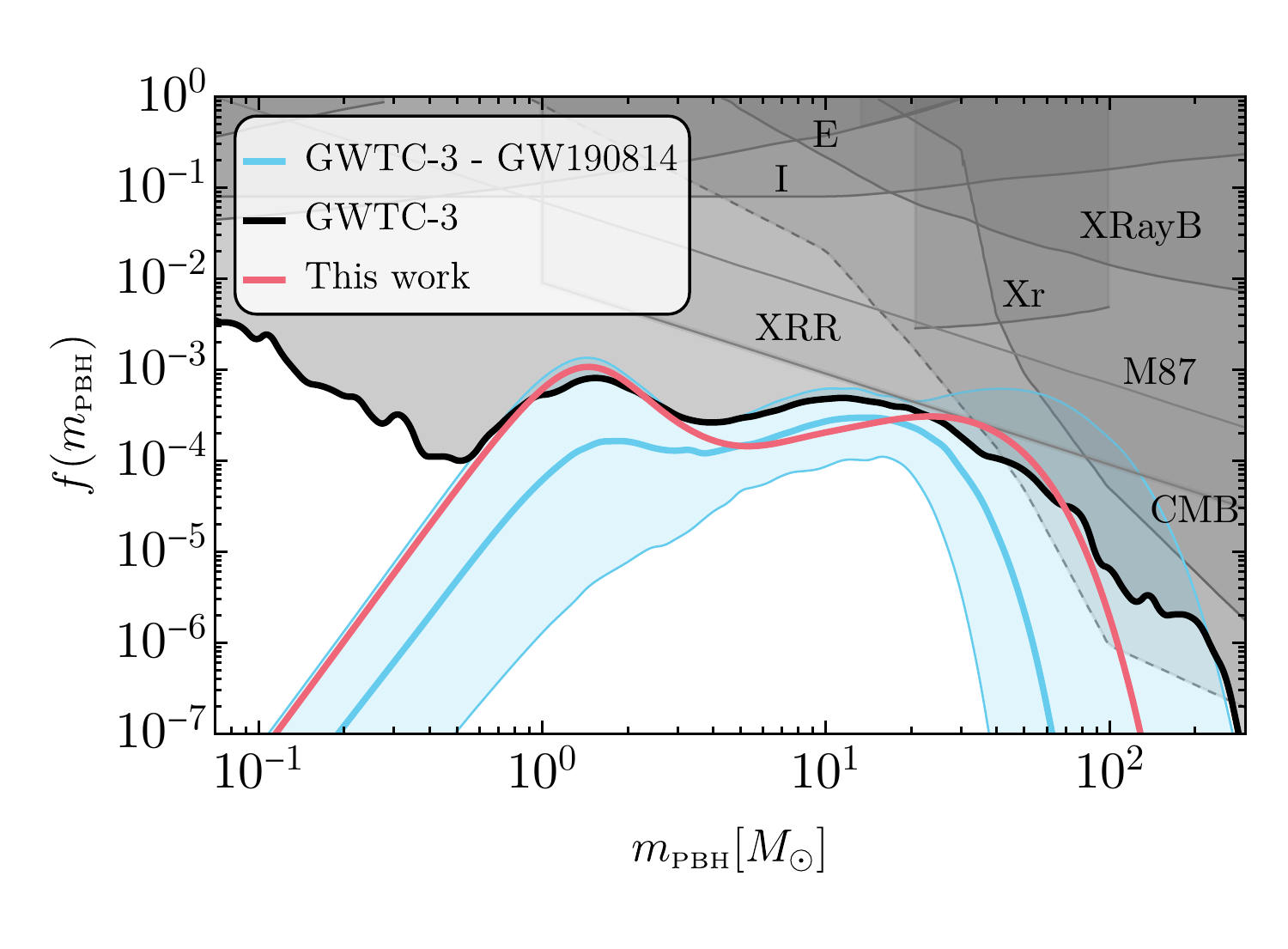}
\includegraphics[width=0.49\textwidth]{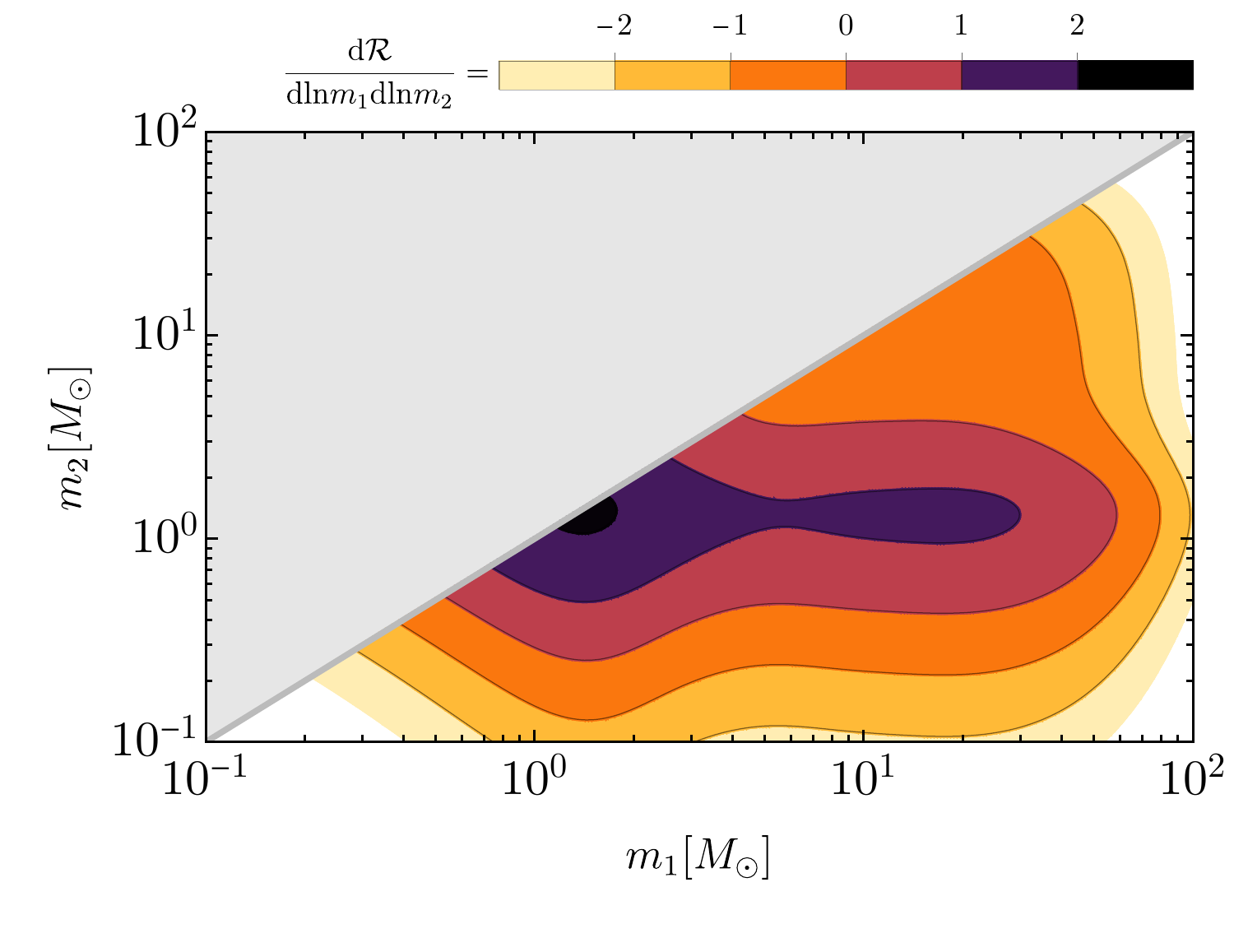}
	\caption{
	{\it Left:}
	Constraints on the PBH mass distribution derived in~\cite{Franciolini:2022tfm} and compared to existing ones~\cite{Carr:2020gox}.
	The thick black line delineates the GWTC--3 upper bound (90\% C.I.) on PBH assuming a dominant contribution from ABH binaries and null detection of subsolar mergers. 
	The cyan band shows the portion of the posterior where GW190814 is interpreted as a PBH binary (see Ref.~\cite{Franciolini:2022tfm} for more details). 
	The red line is the mass function we consider in this work to derive optimistic detection prospects at the ET. 
	%
 {\it Right:}
	Differential merger rate (in units of \si{\per\cubic\giga\parsec\per\year}) as a function of primary and secondary mass we assume in this work. 
	}
\label{fig:PBH_constraints}
\end{figure*}

 The suppression factor $S<1$  accounts for environmental effects in both the early-- and late--time Universe. We can separately define each contribution as
\begin{equation}
S \equiv S_\text{\tiny early}(M, f_\PBH, \psi) \times S_\text{\tiny late}(f_\PBH,z)\,.    
\end{equation}
An analytic expression for $S$ can be found in Ref.~\cite{Hutsi:2020sol}, which we report here for completeness. 
In the early Universe, suppression results as a consequence of interactions between PBH binaries and both the surrounding dark matter inhomogeneities, as well as neighboring PBHs at high redshift~\cite{Eroshenko:2016hmn,Ali-Haimoud:2017rtz,Raidal:2018bbj,Liu:2018ess}. 
This factor takes the form
\begin{align}
S_\text{\tiny early}& 
	\thickapprox 
	1.42 \llp \frac{\langle m^2 \rangle/\langle m\rangle^2}{\bar N(y) +C} + \frac{\sigma ^2_\text{\tiny M}}{f^2_\PBH}\rrp ^{\nicefrac{-21}{74}} 
	\exp \llp -  \bar N\rrp\,, 
    \label{S1}
\end{align}
with
\begin{align}
\bar N
\equiv 
\frac{M}{\langle m \rangle }
\lp \frac{f_\PBH}{f_\PBH+ \sigma_\text{\tiny M}} \rp \,,
\end{align}
and the rescaled variance of matter density perturbations takes the value $\sigma_\text{\tiny M}\simeq 0.004$.
We also introduced the first and second momenta of the PBH average mass distribution, i.e. 
$\langle m \rangle$ and $\langle m^2 \rangle$, respectively.
In Eq.~\eqref{S1}, the constant factor $C$ is defined as (see Eq.~(A.5) of Ref.~\cite{Hutsi:2020sol})
\begin{align}
    C 
     &= 
     \frac{ }{} \frac{}{}
    \frac{f_\PBH^{2} \times  \langle m^2\rangle / (\sigma_\text{\tiny M} \times \langle m\rangle) ^2}
    { \left[ \frac{\Gamma(29/37)} {\sqrt{\pi}} U\left(\frac{21}{74},\frac{1}{2},\frac{5f_\PBH^{2}}{6 \sigma_\text{\tiny M}^{2}}\right) \right]^{\nicefrac{-74}{21}} - 1 }\,,
\end{align}
where $\Gamma(x)$ is the Euler Gamma function and $U(a,b,z)$ denotes the confluent hypergeometric function.

In the late Universe, multiple encounters with other PBHs that populate small clusters formed from the initial Poisson conditions lead to a thermalisation of the eccentricity distribution, which enhances the merger time and effectively reduces the late--time universe merger rate~\cite{Jedamzik:2020ypm,Young:2020scc,Jedamzik:2020omx,Trashorras:2020mwn,Tkachev:2020uin}. By accounting for the fraction of binaries that avoids dense enough clusters and are not disrupted, one can write down this additional suppression factor as~\cite{Vaskonen:2019jpv,DeLuca:2020jug,Hutsi:2020sol,link}
\begin{align}
S_\text{\tiny late} (x) & \thickapprox \text{min} \llp 1,\, \num{9.6e-3} x ^{-0.65} \exp \lp 0.03 \ln^2 x \rp  \rrp\,,
\end{align}
where we introduced the variable $x \equiv (t(z)/t_0)^{0.44} f_\PBH$.
Notice also that, for $f_\PBH \lesssim 0.003$, one always finds \mbox{$S_\text{\tiny late}\simeq 1$}, i.e. the suppression of the merger rate due to disruption inside PBH clusters is negligible. This is also supported by the results obtained through cosmological $N$--body simulations finding that PBHs are essentially isolated when their abundance is small enough~\cite{Inman:2019wvr}.

\subsection{GWTC--3 upper bound}\label{UpperPop}

Following Ref.~\cite{Franciolini:2022tfm}, we assume the PBH population is generated by an enhanced curvature power spectrum in between two scales, see Sec.~II.E of~\cite{Franciolini:2022tfm} for more details. 
The population is derived including a state--of--the--art modelling of the QCD era, and its impact on the threshold for PBH collapse and the critical mass spectrum (see also~\cite{Jedamzik:1996mr,Byrnes:2018clq,Muscoinprep,Escriva:2022bwe}).
Only four hyper--parameters fix the PBH population:
the PBH abundance $f_\text{\tiny PBH}$, the power spectrum tilt $n_s$ (not related to the one characterising the larger CMB scales), the minimum $M_\text{\tiny S}$ and maximum $M_\text{\tiny L}$  PBH masses, linked to the smallest and largest horizon masses bracketing the PBH formation epoch. 
In particular, we use
$f_\PBH = 10^{-2.8}$, $n_s = 0.84$, 
	$M_\text{\tiny S} = 10^{-1.6}\,\si{\Msun}$ and $M_\text{\tiny L} = \SI{e3}{\Msun}$,
 which corresponds to a representative upper bound on the PBH population given the currently available GWTC--3 dataset.

The red line in Fig.~\ref{fig:PBH_constraints} (left panel)  is the mass function that we consider in this work, alongside the most stringent constraint on this window. 
With this choice, the local rate density is 
\begin{equation}
    {\cal R}_\PBH (z = 0) =  \SI{89}{\per\cubic\giga\parsec\per\year}\,,
\end{equation}
integrated over all masses in the range $m_i \in [10^{-2},10^3]~\si{\Msun}$.	
Notice, in particular, that this number is
dominated by the contribution of light events 
and it is smaller than 
what is estimated by the LVK Collaboration in the ``Full'' mass range~\cite{2021arXiv211103634T}, as PBH can only  
explain a small fraction of current detected events~\cite{Franciolini:2021tla,Franciolini:2022tfm}.
The intrinsic distribution of mergers in both primary and secondary mass is shown in the right panel of Fig.~\ref{fig:PBH_constraints}.

The total number of mergers per year in the redshift window 
$z \in [0,100]$ is
\begin{equation}\label{integrandz}
    N^{z \in [0,100]}_\PBH =
    \int_0^{100} \d z 
    \frac{\d V_c}{\d z} 
    \frac{{\cal R}_\PBH (z) }{1+z} 
    = \SI[per-mode=symbol]{1.92e6}{\per\year}\,,
\end{equation}
which is the number of sources we analyse.
We do not consider higher redshift sources, even though they exist, as above $z = 100$ the selection bias reduces to zero the observable mergers in all configurations we considered. 
The sources are distributed uniformly across the sky and we also assume uniform distributions for the inclination angle, polarisation angle, phase at coalescence, and merger time.

Consistently with the above assumption of negligible effect of accretion on PBH masses adopted in Ref.~\cite{Franciolini:2022tfm}, the spin magnitudes are drawn assuming small PBH accretion efficiency and 
adopting the accretion model described in detail in Refs.~\cite{DeLuca:2020bjf,DeLuca:2020fpg,DeLuca:2020qqa}.
For definiteness, we assume a cut--off redshift $z_\text{\tiny cut--off} \approx 23$ 
while we assume uniform and independent distributions for the spin orientations, as discussed in Sec.~\ref{PBHmodel_MassSpins}.

\subsection{Population detection prospects at ET}
We perform the analysis on the described population for all the ET configurations considered in this work, also in combination with one or two Cosmic Explorer (CE) detectors for the full HFLF-Cryo ASD, assuming a 40~km detector in the former case and a 40~km plus a 20~km detector in the latter. As a comparison, we further perform the analysis for the LVKI network using the most optimistic sensitivity curves currently available for the O5 run.

{
\renewcommand{\arraystretch}{1.4}
\setlength{\tabcolsep}{10pt}
\begin{table*}[t!]
\begin{tabularx}{2.\columnwidth}{||X||c||c|c|c||c|c||}
\hline\hline
 &
        $N^\text{\tiny tot}$ &
        $N^\text{\tiny SS}$ &
        $N^{z>10}$ &
        $N^{z>30}$ &
        $N^\text{\tiny LMG}$ &
        $N^\text{\tiny UMG}$ 
\\
\hline\hline
Intrinsic population &
        \num{1920000} &
        \num{708487} &
        \num{1400384} &
        \num{795904} &
        \num{300220} &
        \num{7774} 
\\
\hline \hline
Configuration & 
        $N^\text{\tiny tot}_\text{\tiny det}$ &
        $N_\text{\tiny det}^\text{\tiny SS}$ &
        $N_\text{\tiny det}^{z>10}$ &
        $N_\text{\tiny det}^{z>30}$ &
        $N_\text{\tiny det}^\text{\tiny LMG}$ &
        $N_\text{\tiny det}^\text{\tiny UMG}$ 
\\
       \hline\hline
        $\bigtriangleup$-10km-HFLF-Cryo& \num{13347} & \num{1650} & \num{336} & \num{17} & \num{2638} & \num{235} \\
        $\bigtriangleup$-15km-HFLF-Cryo& \num{30912} & \num{4281} & \num{1099} & \num{91} & \num{6443} & \num{376} \\
        2L-15km-\ang{45}-HFLF-Cryo& \num{24900} & \num{3345} & \num{824} & \num{66} & \num{5132} & \num{332} \\
        2L-15km-\ang{0}-HFLF-Cryo& \num{26585} & \num{3580} & \num{940} & \num{65} & \num{5517} & \num{356} \\
        2L-20km-\ang{45}-HFLF-Cryo& \num{35524} & \num{5206} & \num{1434} & \num{140} & \num{7550} & \num{374} \\
        2L-20km-\ang{0}-HFLF-Cryo& \num{45650} & \num{6745} & \num{1962} & \num{187} & \num{9809} & \num{465} \\
        1L-20km-HFLF-Cryo & \num{22852} & \num{3019} & \num{698} & \num{37} & \num{4656} & \num{310} \\
        \hline\hline
        $\bigtriangleup$-10km-HF& \num{4804} & \num{553} & \num{4} & \num{0} & \num{1023} & \num{74} \\
        $\bigtriangleup$-15km-HF& \num{12739} & \num{1660} & \num{106} & \num{0} & \num{2747} & \num{130} \\
        2L-15km-\ang{45}-HF & \num{10714} & \num{1379} & \num{76} & \num{0} & \num{2239} & \num{113} \\
        2L-15km-\ang{0}-HF& \num{10762} & \num{1374} & \num{79} & \num{0} & \num{2297} & \num{114} \\
        2L-20km-\ang{45}-HF& \num{20704} & \num{2900} & \num{286} & \num{0} & \num{4574} & \num{175} \\
        2L-20km-\ang{0}-HF& \num{15951} & \num{2211} & \num{210} & \num{0} & \num{3515} & \num{146} \\
        1L-20km-HF & \num{7610} & \num{987} & \num{51} & \num{0} & \num{1633} & \num{84} \\
        \hline\hline
        CE-40km & \num{41650} & \num{6250} & \num{2050} & \num{142} & \num{8732} & \num{378} \\
        (CE-40km)-(CE-20km) & \num{55832} & \num{8702} & \num{2995} & \num{223} & \num{11897} & \num{469} \\
        \hline\hline
        (CE-40km)-($\bigtriangleup$-10km-HFLF-Cryo) & \num{56964} & \num{8668} & \num{2876} & \num{226} & \num{12108} & \num{510} \\
        (CE-40km)-(2L-15km-\ang{45}-HFLF-Cryo) & \num{68524} & \num{10669} & \num{3604} & \num{331} & \num{14713} & \num{590} \\
        (CE-40km)-(2L-15km-\ang{0}-HFLF-Cryo) & \num{65245} & \num{10127} & \num{3366} & \num{294} & \num{13990} & \num{563} \\
        (CE-40km)-(CE-20km)-($\bigtriangleup$-10km-HFLF-Cryo) & \num{71414} & \num{11300} & \num{3945} & \num{332} & \num{15430} & \num{589} \\
        (CE-40km)-(CE-20km)-(2L-15km-\ang{45}-HFLF-Cryo) & \num{82812} & \num{13297} & \num{4757} & \num{443} & \num{17975} & \num{663} \\
        (CE-40km)-(CE-20km)-(2L-15km-\ang{0}-HFLF-Cryo) & \num{77974} & \num{12444} & \num{4379} & \num{396} & \num{16827} & \num{621} \\
        \hline\hline
        LVKI-O5 & \num{49} & \num{7} & \num{0} & \num{0} & \num{10} & \num{1} \\
        \hline\hline
\end{tabularx}
\caption{
{\it Top row:} Intrinsic number of PBH merger per year predicted by the population we assume in this section. 
{\it Bottom rows:}
Number of detected events (${\rm SNR}\geq12$) per year for all the ET configurations considered in this work.
In particular, in the second column we report the total number of detections, in the third the number of detections in the subsolar range (i.e. having at least the lightest object in the binary in the subsolar window $m_2\leq\SI{1}{\Msun}$), in the fourth and fifth the number of detections at high redshift, $z\geq10$ and $z\geq30$, respectively, in the sixth column the number of detected events having at least one of the masses in the so--called lower mass gap 
(i.e. $m_1\in[2.5,5]~\si{\Msun}$ and/or $m_2\in[2.5,5]~\si{\Msun}$) 
and in the last columns the number of observed events with at least one of the masses in the so--called upper mass gap (i.e. having at least the heavier object in the binary heavier than $m_1\geq\SI{50}{\Msun}$).}
\label{tab:allConf}
\end{table*}
}

\begin{figure*}[!ht]
	\centering
	\includegraphics[width=0.68\textwidth]{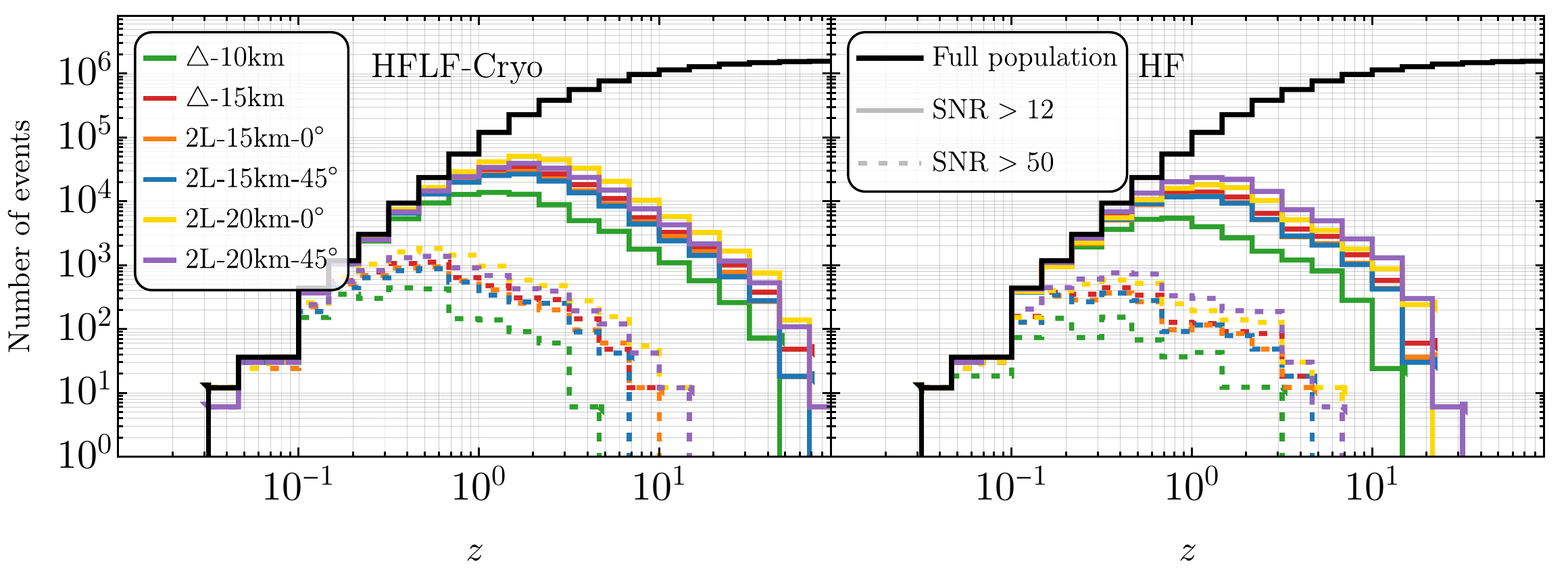}
 \includegraphics[width=1.0\textwidth]{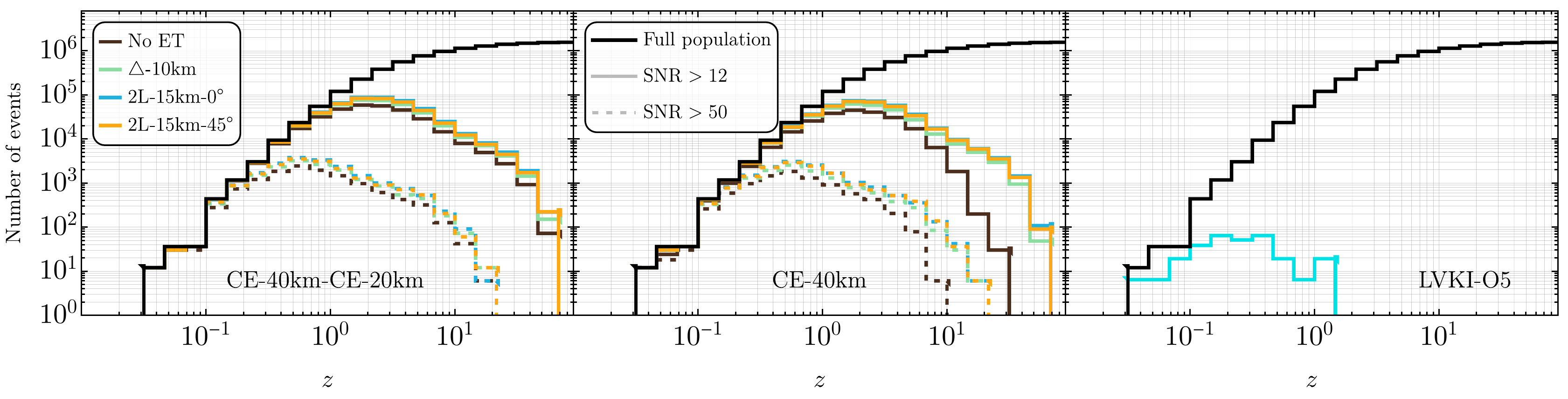}
	\caption{
Probability density function of detected events (with SNR$>12$ and SNR$>50$) for 
the population assumed in Sec.~\ref{UpperPop}. 
The black solid line indicates the entire population of mergers per year. 
The number of detection per bin is monotonically increasing as we choose a binning equally spaced in log--scale. 
In practice, the event distribution in redshift follows 
the integrand of Eq.~\eqref{integrandz}, 
which peaks at redshift $z\approx \order{10}$ because of the comoving volume factor. 
	}
\label{fig:popSNR_vsz}
\end{figure*}

\begin{figure*}[!ht]
	\centering
	\includegraphics[width=0.68\textwidth]{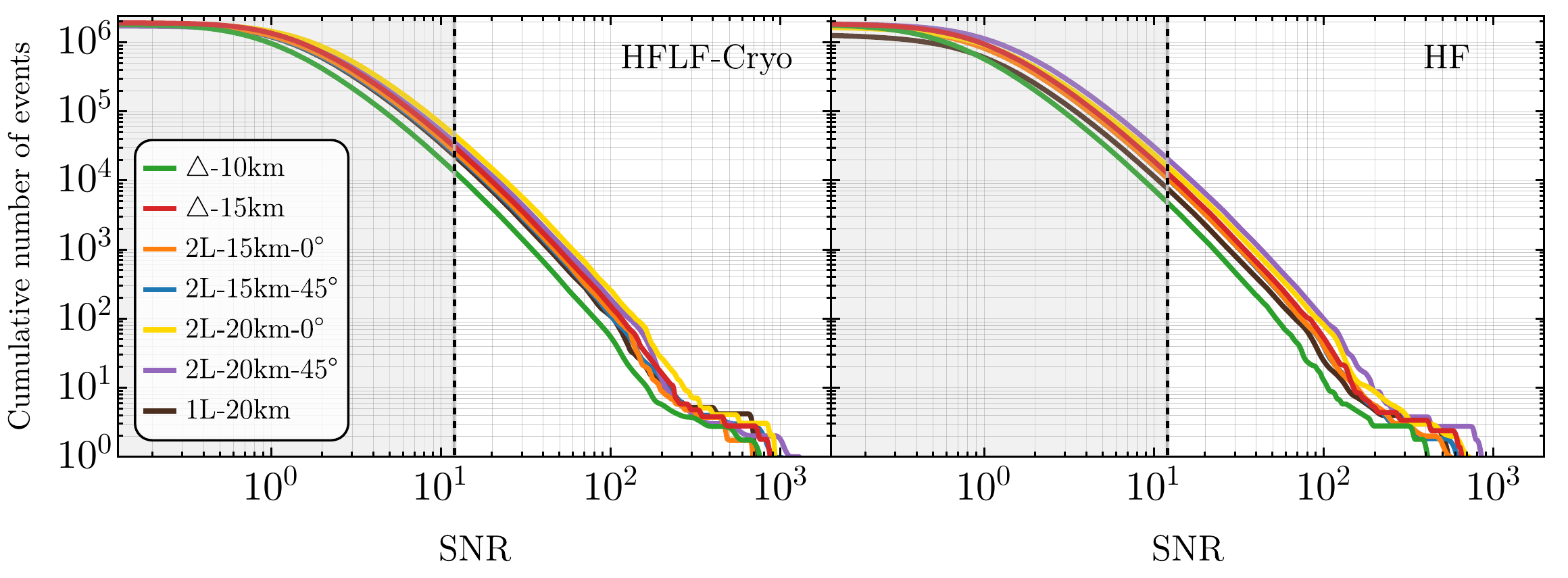}
 \includegraphics[width=1\textwidth]{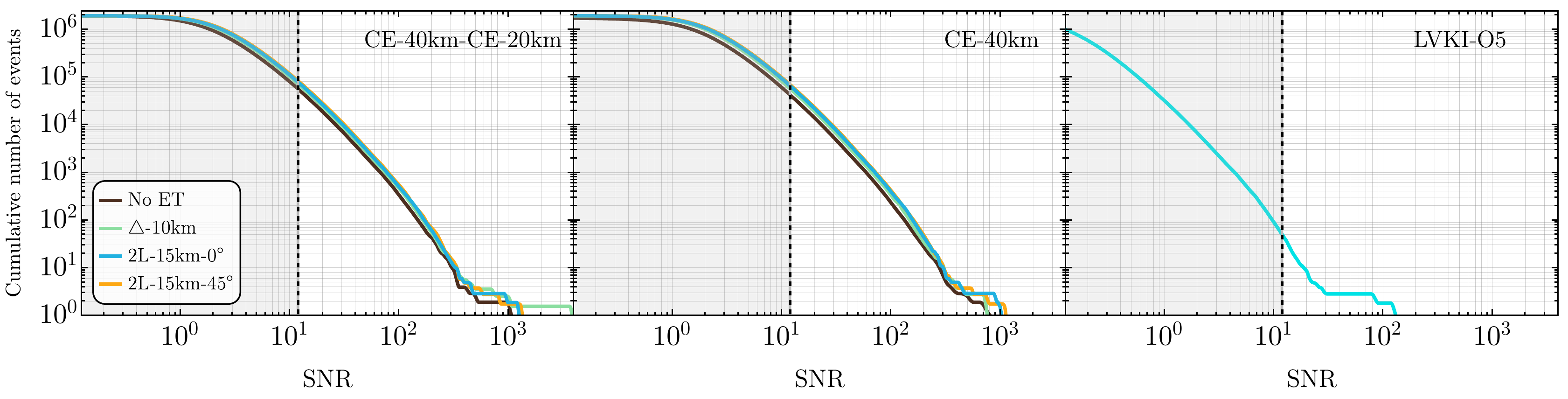}	
	\caption{
	Cumulative distribution of the SNRs for the simulated PBH population as observed by the different detector configurations. 
 The cumulative is inverted to actually show the number of events with SNR higher than a given value, and we shade the region ${\rm SNR}<12$, which is below the assumed detection threshold. 
 Top panel: configurations featuring only the ET detector geometries considered with their HFLF-Cryo sensitivity (left panel) and HF sensitivity (right panel).
 Bottom panel: network of detectors which features either two CE (with and without ET) and, on the rightmost side, the optimistic O5 LVKI detector performance. 
	}
\label{fig:popSNR}
\end{figure*}

The analysis is performed using the \textsc{IMRPhenomXPHM} waveform model for the full set of 15 parameters 
\begin{equation}
    \{{\cal M}_c, \eta, d_L, \theta, \phi, \theta_{JN}, \psi, t_c, \Phi_c, \chi_{1}, \chi_{2}, \theta_{s,1}, \theta_{s,2}, \phi_{JL}, \phi_{1,2}\}\,,
\end{equation}
where $\chi_{1}$ and $\chi_{2}$ denote the spin magnitudes of the two objects, $\theta_{JN}$ is the angle between the total angular momentum vector of the binary and the line of sight, $\theta_{s,1}$ and $\theta_{s,2}$ are the tilts of the two spins, $\phi_{JL}$ is the  azimuthal angle of the orbital angular momentum relative to the total angular momentum, and $\phi_{1,2}$ the difference in azimuthal angle of the two spins. For each event we compute the SNR at the various detector configuration, while the parameter estimation is performed only for the events having ${\rm SNR}\geq12$.\footnote{We further discard a small fraction of events having an ill--conditioned Fisher matrix, which is found never to exceed 1\%. See Ref.~\cite{Iacovelli:2022bbs} for a discussion of the methodology.} To have a more realistic simulation, we assume an uncorrelated 85\% duty cycle for 3G detectors (this is applied to each arm for the triangle) and a 70\% duty cycle for 2G detectors during O5.

In Tab.~\ref{tab:allConf}, we summarise the results for the number of detected events, imposing a conservative detection threshold ${\rm SNR}\geq12$, which is the same adopted for the parameter estimation. In particular, in the second column, we report the results for the total number of observed events.
\footnote{Real world search pipelines are based on more advanced statistics, such as the false-alarm-rate. However, this requires a detailed understanding of the level of non-gaussian noise in the detector, which can only be obtained after a detector has been built and commissioned.  Nevertheless, we expect the SNR criterion to allow for a consistent comparison between configurations, as well as a reliable estimate of the 
performance of the detectors. }

Furthermore, following the logic described in Ref.~\cite{Franciolini:2022tfm}, we report the number of detections per year falling in ``special'' mass ranges, which would somehow challenge vanilla astrophysical scenarios.  These are of particular interest since the astrophysical mechanisms producing BH with such masses are poorly understood, and PBHs offer a natural alternative. 
In particular, in the third column we restrict to the systems having at least one of the masses lighter than \SI{1}{\Msun} (with our convention $m_1\geq m_2$, this condition is equivalent to $m_2\leq\SI{1}{\Msun}$). In the fourth (fifth) column instead, we report the number of observations of events at redshift $z\geq10$ ($z\geq30$). Finally, in the sixth column, we report the number of observed sources having at least one of the objects with a mass falling in the so--called lower mass gap (LMG), i.e. in the range $[2.5,5]~\si{\Msun}$, while in the seventh column the ones having at least one object with a mass falling in the upper mass gap (UMG) i.e. above \SI{50}{\Msun} (this condition is equivalent to $m_1\geq\SI{50}{\Msun}$). 
In the top rows of Tab.~\ref{tab:allConf}, we also report the number of mergers produced by the intrinsic population per year (i.e. before the selection bias of the detector is applied). This helps gain intuition on how restrictive is the limited sensitivity of the detector within the various range of the mass-distance parameter space.

Already from these results a number of conclusions can be drawn: comparing the HFLF-Cryo and HF results we see the fundamental importance of the LF instrument in the ET design, which enables to observe binaries up to extremely high redshifts, one of the strong signatures of PBHs, while also more than doubling the number of overall detections. We further notice the overall better results obtained with the 2L-15km  configuration as compared to the 10~km triangular geometry. Given that the adopted population predicts a high number of binaries with low masses (around \SI{1}{\Msun} and \SI{2}{\Msun} as can be seen from Fig.~\ref{fig:PBH_constraints}), we also appreciate the performance of the CE detectors, which are more sensitive than ET for such events (see e.g. Fig.~4 of Ref.~\cite{Iacovelli:2022bbs}), but still notice the significant improvement brought by ET, showing the potential of the full 3G network. Finally, comparing the results obtained for 3G detectors to the ones for the LVKI~O5 network, we see the huge leap achievable thanks to ET and CE.

The overall conclusion we draw from these findings is that 3G detectors will be able to discover a PBH population which may be undetected by current experiments. 
Both the subsolar and high-redshift smoking-gun signatures
would provide unique probes of the existence of such population, 
with the latter being accessible only if the HFLF-Cryo ET configuration is achieved. 
In both mass gaps, the number of detectable events can reach multiple thousands, which would further allow to perform dedicated population analyses aimed at distinguishing these contributions from astrophysical contaminats in those regions. 
In case such events were not discovered, ET would set much more stringent bounds on the abundance of PBHs in the stellar mass range than what can optimistically achieved with current LVK technology \cite{DeLuca:2021hde}.

In Fig.~\ref{fig:popSNR_vsz} we show the distribution of mergers as a function of redshift  $z$. 
Due to the small rates, large search volumes are needed in order to obtain a sizeable number of events. This means that most of the detectable mergers fall in the redshift range between $z\approx \order{0.1}$ and $z\approx \order{40}$, with the latter boundary of this range which strongly depends on the specific design considered. 
While the black line shows the full population for reference, the colored solid and dashed lines correspond to the histogram of events detected with SNR larger than 12 or 50, respectively.  
This plot graphically shows that the HF version of the design is not able to reach mergers beyond redshift $\order{30}$, not even in the largest (2L-20km) implementation, thus missing entirely the high redshift (smoking--gun) window. 
On the other hand, the difference between HF and HFLF-Cryo is reduced in the bulk of events, as similar numbers are obtained below redshifts of a few, where all designs have an outstanding performance. 

Anyway, a note of caution is needed here.
The sole fact that a merger happening at $z\gtrsim30$ is detectable does not assure a good reconstruction of its distance. 
As already discussed in the preceding section, 
individual merger detections require 
sufficient precision on the source redshift  in order to prove the primordial nature of the source (see Refs.~\cite{Ng:2021sqn, Ng:2022vbz} for an in--depth analysis of this parameter estimation uncertainties based on Bayesian PE, and \cite{Mancarella:2023ehn} for figures of merit describing the attainable lower confidence level on the redshift). 
One could also exploit
a population--based inference of the high--redshift merger rate. 
In such a way, multiple event detection can be combined, thus enhancing the statistics and reducing the overall uncertainties
\cite{Ng:2022agi,Martinelli:2022elq}.
In any case, this simplified comparison solidly shows the importance of the cryogenic instrument in order to access the high redshift smoking-gun signature.

\begin{figure*}[!t]
\vspace{-1.cm}
	\centering
	\includegraphics[width=0.75\textwidth]{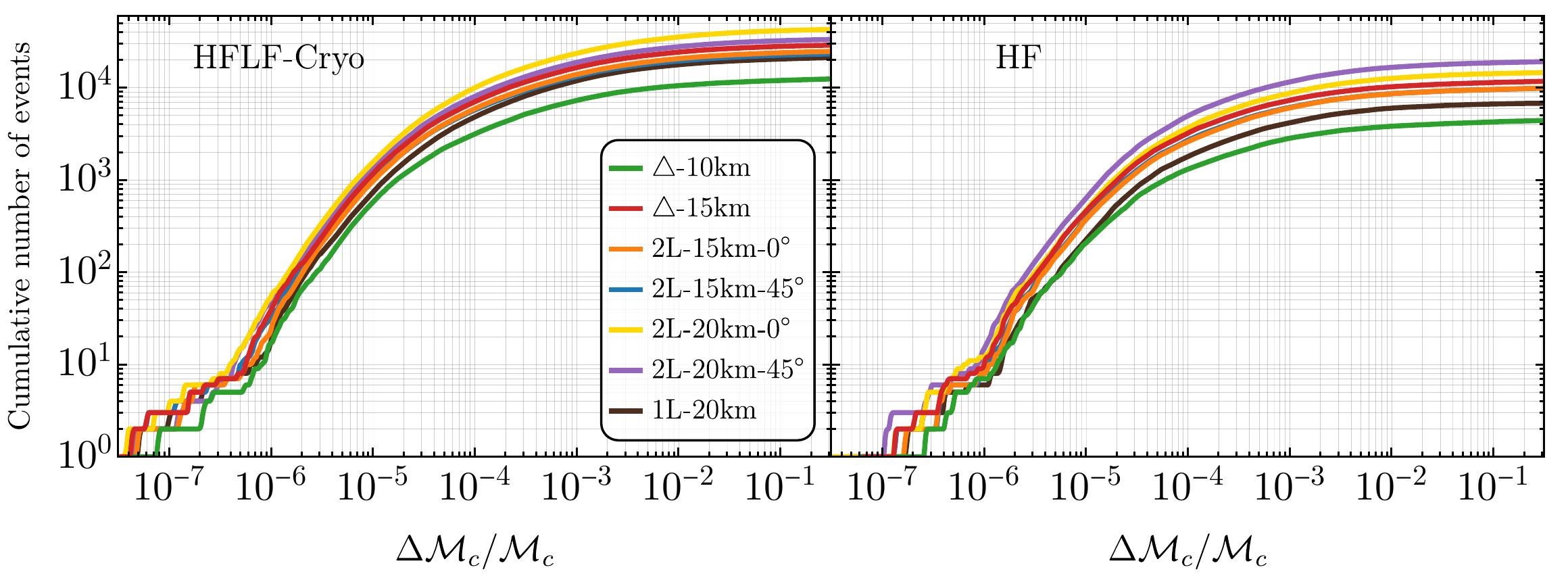}
		\includegraphics[width=0.75\textwidth]{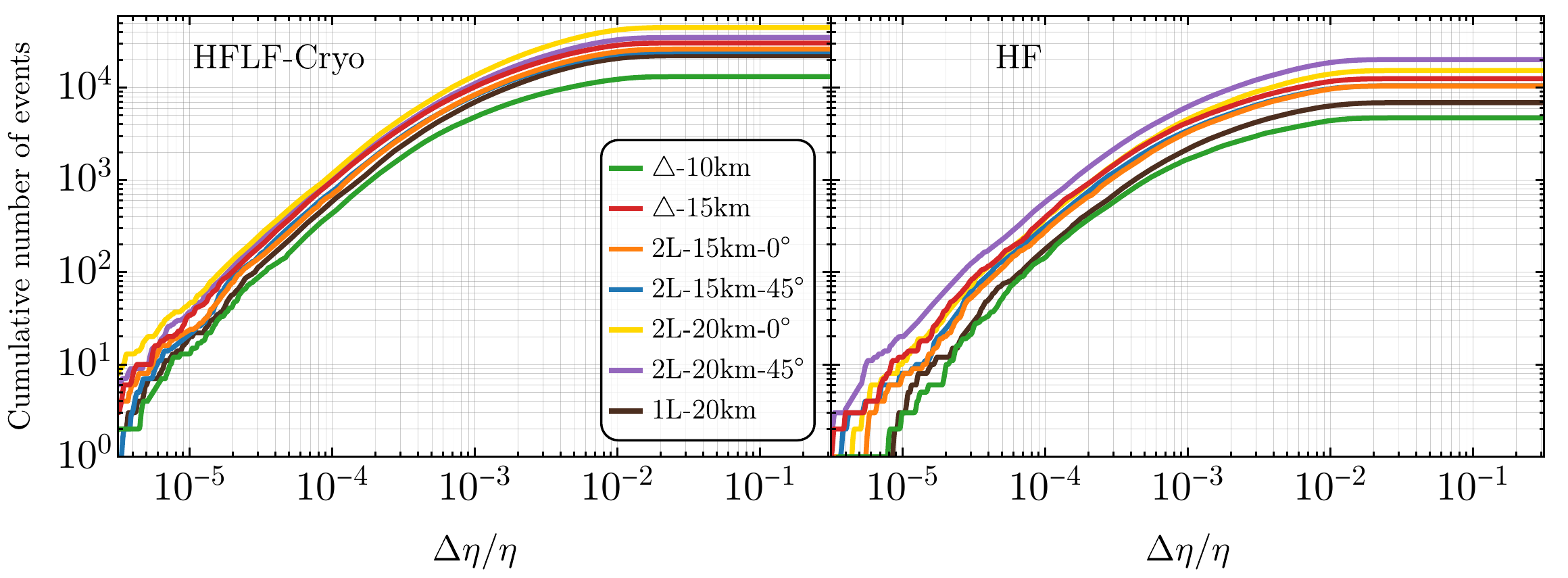}
		\includegraphics[width=0.75\textwidth]{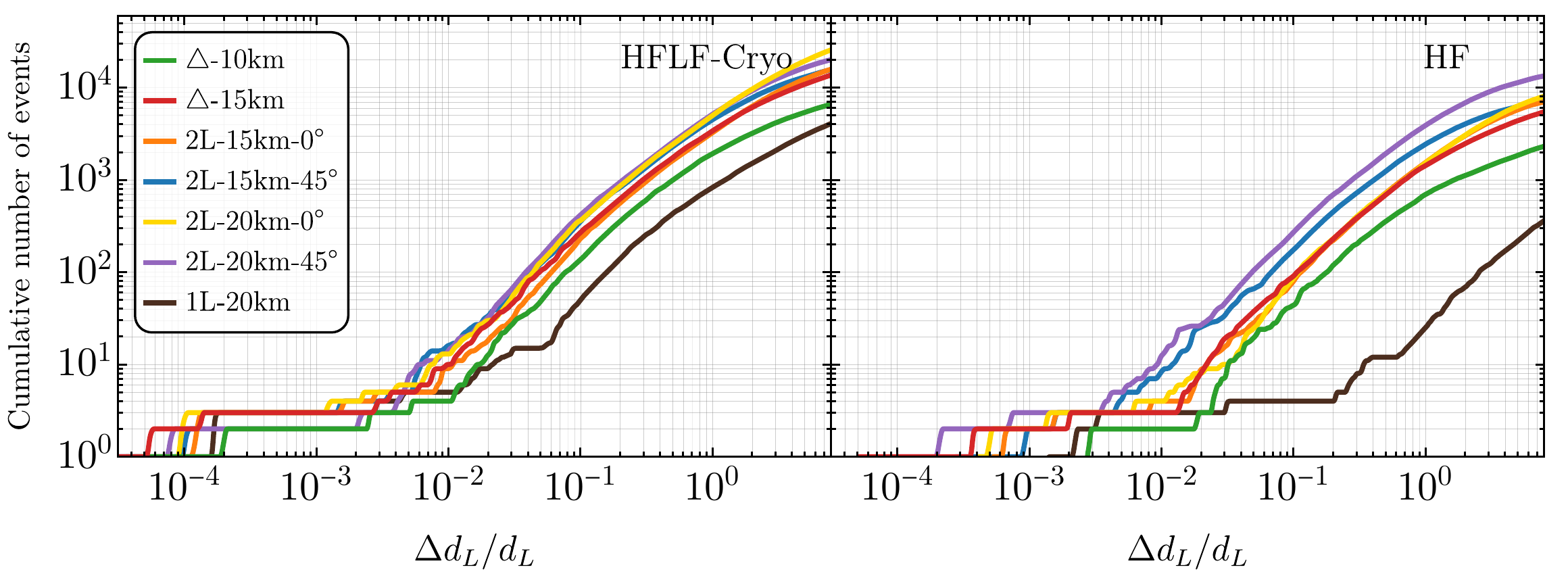}
\includegraphics[width=0.75\textwidth]{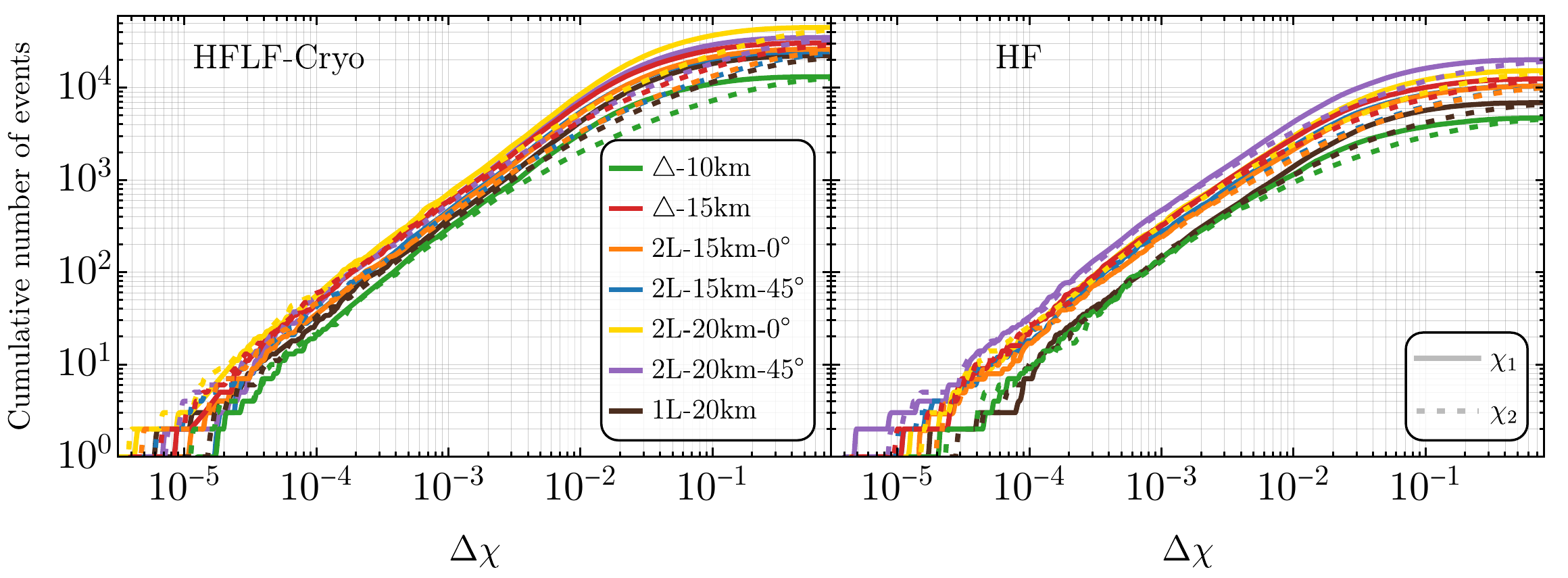}		
	\caption{
	Cumulative distribution of parameter estimation errors for some selected parameters obtained with the Fisher matrix approach for the simulated PBH population as observed by the different detector geometries considered with their HFLF-Cryo sensitivity (left panel) and HF sensitivity (right panel). In particular, in the the top row we report the relative errors attainable on the detector--frame chirp mass, in the second row for the symmetric mass ratio, and in the third row on the luminosity distance, while in the bottom row we report the absolute errors attainable on the adimensional spin magnitude of the primary (solid line) and secondary (dashed line) object. 
	}
\label{fig:popDists}
\end{figure*}

\begin{figure*}[!t]
	\centering
	\includegraphics[width=1\textwidth]{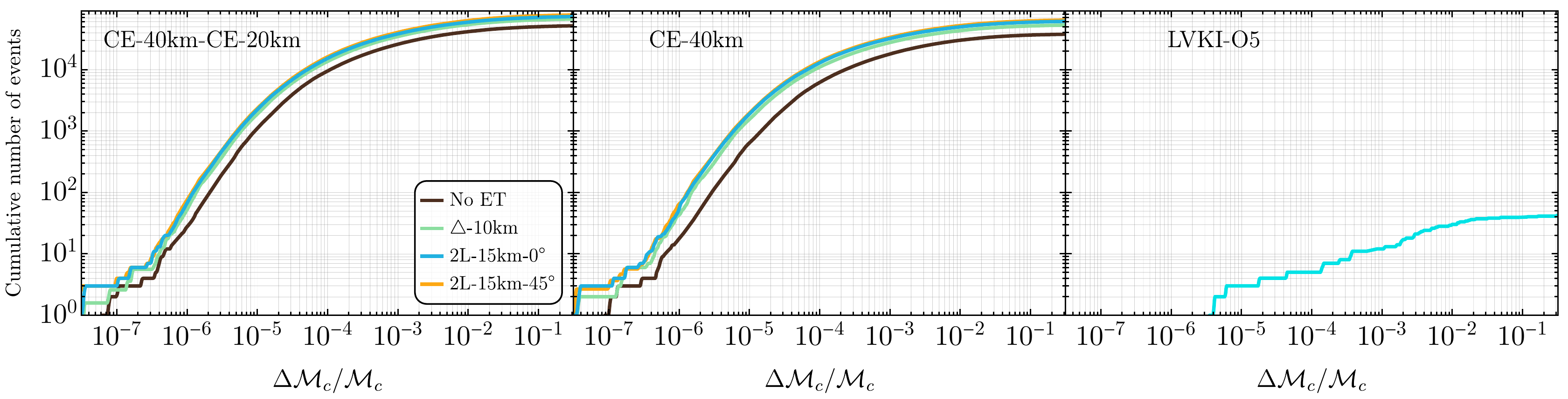}
		\includegraphics[width=1\textwidth]{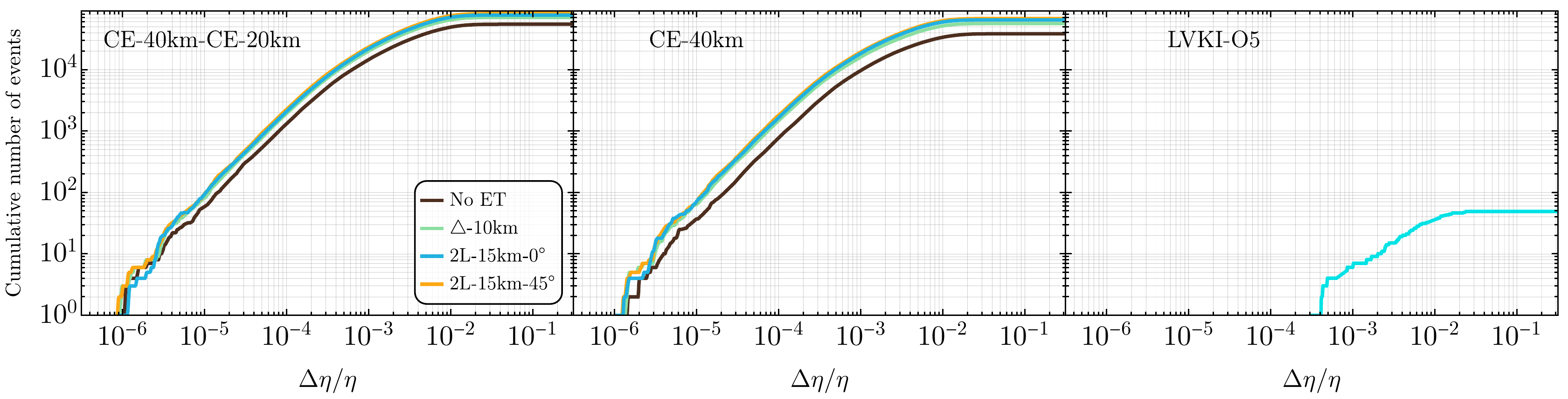}
		\includegraphics[width=1\textwidth]{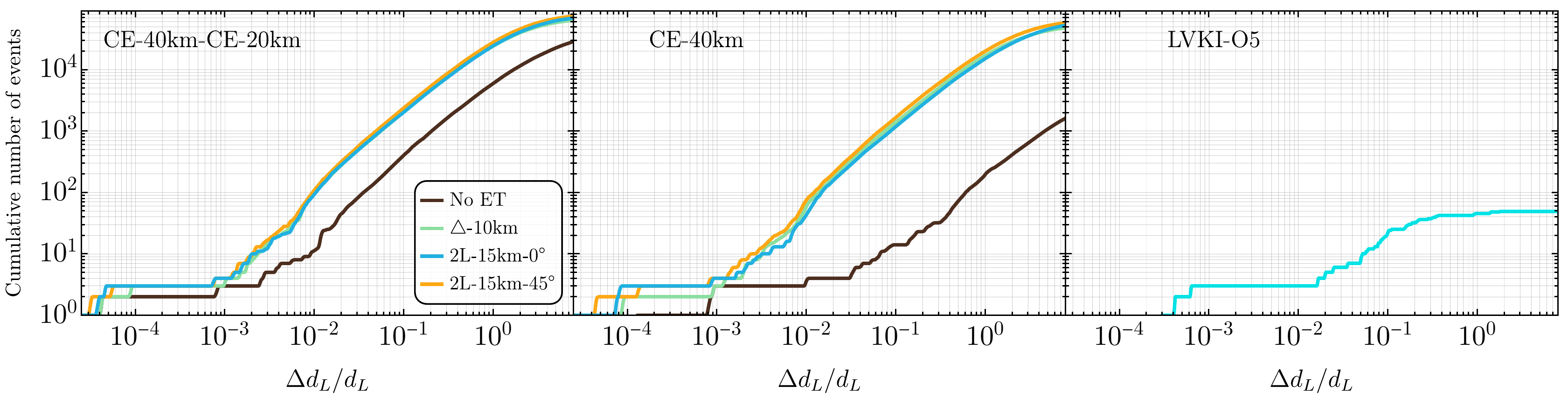}
\includegraphics[width=1\textwidth]{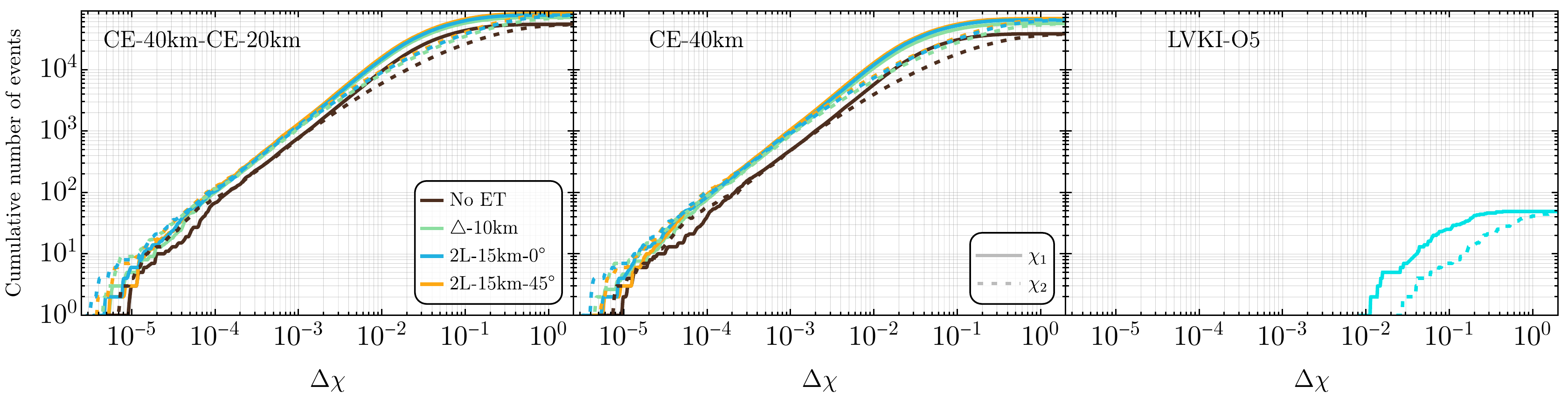}		
	\caption{
	Same as  Fig.~\ref{fig:popDists}, but networks which include:  two CE detectors, one 40~km long and the other 20~km long (left panels) or a single 40~km CE detector (central panel) alone and in a network with the ET both in its 10~km triangular design and in the 15~km 2L designs aligned and misaligned, all with their HFLF-Cryo sensitivity. In the right panels, we report the results obtained for the optimistic version of LVKI network with the best sensitivities representative of the O5 run. 
	}
\label{fig:popDistsCE}
\end{figure*}

We report in Figs.~\ref{fig:popSNR} (top panel) the cumulative distributions of SNRs attainable at ET with all the considered geometries and the two ASDs, while in the bottom panels we show the same results for CE (both as a single 40~km detector and as two instruments of 40~km and 20~km) both alone and in a network with ET, and for the LVKI network during the O5 observing run.

In Figs.~\ref{fig:popDists} and \ref{fig:popDistsCE}, we report the cumulative distribution of parameter estimation errors on the most relevant parameters 
characterising each event in the population of PBHs.

As far as ET alone is concerned, we again notice the overall better results obtained with the 2L misaligned instruments, especially in the HF scenario, both for the measurement of the intrinsic parameters and, in particular, for the luminosity distance reconstruction. Moreover, we again confirm the relevance of the LF instrument, with a number of detections more than two times higher as compared to the HF case, and more events having good parameter reconstruction. Considering ET in a network with CE the differences among the three considered geometries appear instead less pronounced, with the 2L configurations providing anyway better results.


\section{Summary and conclusions}\label{sec:conclusions}

In this paper, we have analysed how different detector designs for the ET would perform on the estimation and discrimination of some key observables of PBH mergers,
as well as their detection prospects for a realistic population of PBH events~\cite{Franciolini:2022tfm}. In particular, we have studied three detector geometries: a triangular detector consisting of three nested instruments and two L--shaped detectors, both parallel and misaligned by \ang{45} among them. For the various geometries we have further considered two different arm lengths and two scenarios: one in which the detectors operate at their full potential and adopt a `xylophone' design, i.e. consist of 2 instruments, one optimised for LF sensitivity and the other for HF sensitivity, and another in which only the HF instrument is operating. For a comprehensive analysis of how other science cases depend on these configurations we refer to Ref.~\cite{Branchesi:2023mws}, where also the results here presented are summarised.  

Overall, we find that, thanks to the improved sensitivity and the wider bandwidth, 
3G detectors are much more likely to detect mergers of PBH binaries as compared to current facilities (mainly due to the extended reach), and have the potential to measure their parameters with sufficient accuracy to distinguish them from the astrophysical BHs. These observables include, in particular: {\it i)} the masses in the sub--solar window (where ABH are not expected to form); {\it ii)} the distance in the high--$z$ window for PBHs of mass around few tens of solar mass (after $z\gtrsim30$ astrophysical BH are not expected to exist);  {\it iii)}  the spins (which are relevant due to the expected mass--spin correlation for PBH). 
This can possibly shed light on the possible  primordial origin of some binaries, even on a single--event basis. 
This is also supported by the comparison
of the results attainable at ET with the ones for the LVKI network during O5 (reported in the context of the population analysis in Figs.~\ref{fig:popDists} and \ref{fig:popDistsCE}), we see how big is the leap from 2G to 3G instruments, giving a hint of how deeply instruments such as ET will enable us to look in the window 2G detectors opened on our universe.

One of the key results emerging from our analyses is the relevance of the low--frequency instrument for PBH observations: losing the LF sensitivity would dramatically reduce the redshift range, as is apparent from Figs.~\ref{fig:horizon} and \ref{fig:distance_error_highz}, at the level of potentially inhibiting the observation of mergers above $z\approx30$, one of the smoking--gun signatures of PBH binaries, and an extremely interesting region of our universe to explore. The LF instrument turns out to be fundamental also for measuring a possible orbital eccentricity (PBH binaries are expected to have almost circular orbits), and enables to detect more than twice the number of events as compared to the HF case looking at the population analysis, also resulting in tighter parameter errors, irrespectively of the geometry. 

We further find the longer detectors to be preferable, as expected, in particular for the observable range and the number of detections, as well as for parameter estimation.
Concerning the detector geometry, we find the 2L configuration with misaligned arms to provide the best option. In particular, it provides a better estimation performance 
on all the parameters and enables to observe a higher number of events as compared to the triangular design [we are here comparing the 2L design with \SI{15}{\kilo\meter} (\SI{20}{\kilo\meter}) arms and the triangular design with \SI{10}{\kilo\meter} (\SI{15}{\kilo\meter}) arms].\footnote{What will be the most appropriate comparison between the arm-length of the triangle and 2L configurations can only be ascertained with a detailed analysis of the relative costs (as well as, possibly, of the different funding schemes), currently under development within the ET collaboration.}
When considering ET in a network with CE, the differences among the various detector geometries are smaller, as can be seen from the population analysis results, with the 2L configurations still being preferred. \\

\let\oldaddcontentsline\addcontentsline
\renewcommand{\addcontentsline}[3]{}
\begin{acknowledgments}
We thank Giancarlo Cella, Valerio De Luca, Gianluca Guida, Frank Ohme and Ed Porter for useful discussions.
The research leading to these results has been conceived and developed within the ET Observational Science Board (OSB).
G.F. and P.P. acknowledge financial support provided under the European
Union's H2020 ERC, Starting Grant agreement no.~DarkGRA--757480 and under the MIUR PRIN programme, and support from the Amaldi Research Center funded by the MIUR program ``Dipartimento di Eccellenza'' (CUP:~B81I18001170001). This work was supported by the EU Horizon 2020 Research and Innovation Programme under the Marie Sklodowska-Curie Grant Agreement No. 101007855. 
The research of F.I. and 
M.~Maggiore is supported by  the  Swiss National Science Foundation, grant 200020$\_$191957, and  by the SwissMap National Center for Competence in Research. F.I. is also supported by the Istituto Svizzero ``Milano Calling'' fellowship.
M.~Mancarella is supported by European Union's H2020 ERC Starting Grant No. 945155-GWmining and Cariplo Foundation Grant No. 2021-0555. 
Computations made use of the Yggdrasil cluster at the University of Geneva.
\end{acknowledgments}
\let\addcontentsline\oldaddcontentsline
\let\oldaddcontentsline\addcontentsline
\renewcommand{\addcontentsline}[3]{}
\bibliography{main}
\let\addcontentsline\oldaddcontentsline
\end{document}